\documentclass[11pt,a4paper]{article}
\usepackage{jheppub}
\usepackage{graphicx}
\usepackage[latin2]{inputenc}
\usepackage{t1enc}
\usepackage{amsmath}
\usepackage{amssymb}
\usepackage{epsfig}
\usepackage{subfigure}
\usepackage{dsfont}
\usepackage{slashed}
\usepackage{multirow}
\usepackage{lscape}
\usepackage{wrapfig}
\usepackage[small,bf]{caption}

\usepackage{setspace}
\newcommand{\be}{\begin{equation}}
\newcommand{\ee}{\end{equation}}
\newcommand{\Tr}{\textmd{Tr}}

\newcommand{\Z}{\mathcal{Z}}
\newcommand{\D}{\mathcal{D}}

\title{The QCD phase diagram for external magnetic fields}

\author[1]{G.~S.~Bali,}
\author[1]{F.~Bruckmann,}
\author[1]{G.~Endr\H{o}di,$^\dagger$}
\author[2,3,4]{Z.~Fodor,}
\author[3]{S.~D.~Katz,}
\author[2,4]{S.~Krieg,}
\author[1]{A.~Sch\"afer,}
\author[2]{K.~K.~Szab\'o}

\affiliation[1]{Institute for Theoretical Physics, Universit\"at Regensburg, D-93040 Regensburg, Germany.}
\affiliation[2]{Department of Physics, Bergische Universit\"at Wuppertal, D-42119, Germany.}
\affiliation[3]{Institute for Theoretical Physics, E\"otv\"os University, H-1117 Budapest, Hungary.}
\affiliation[4]{J\"ulich Supercomputing Centre, Forschungszentrum J\"ulich, D-52425 J\"ulich, Germany. }

\emailAdd{gergely.endrodi@physik.uni-regensburg.de}

\abstract{
The effect of an external (electro)magnetic field on the finite temperature transition of QCD is studied. We generate configurations at various values of the quantized magnetic flux with $N_f=2+1$ flavors of stout smeared staggered quarks, with physical masses. Thermodynamic observables including the chiral condensate and susceptibility, and the strange quark number susceptibility are measured as functions of the field strength. We perform the renormalization of the studied observables and extrapolate the results to the continuum limit using $N_t=6,8$ and $10$ lattices. We also check for finite volume effects using various lattice volumes. We find from all of our observables that the transition temperature $T_c$ significantly decreases with increasing magnetic field. This is in conflict with various model calculations that predict an increasing $T_c(B)$. From a finite volume scaling analysis we find that the analytic crossover that is present at $B=0$ persists up to our largest magnetic fields $eB \approx 1 \textmd{ GeV}^2$, and that the transition strength increases mildly up to this $eB\approx1 \textmd{ GeV}^2$.
}


\arxivnumber{1111.4956}

\begin{document}

\maketitle

%\newpage

\section{Introduction}
\label{sec:intro}

The properties of QCD in strong magnetic\footnote{Throughout the paper `magnetic' refers to electromagnetic i.e. not chromomagnetic.} fields are relevant for at least three important physical situations. First, cosmological models suggest that extremely strong magnetic fields ($\sqrt{eB}\sim2$ GeV) could be produced during the electroweak phase transition of the early universe. This effect might also have an impact on subsequent strong interaction processes~\cite{Vachaspati:1991nm}. Second, large magnetic fields ($\sqrt{eB}\sim 1 $ MeV) are present in the interior of dense neutron stars called magnetars~\cite{Duncan:1992hi}. Finally, in a noncentral heavy ion collision the spectators -- being two beams of positive charges moving in opposite directions -- also create an intense magnetic field which, depending on the centrality and the beam momentum, reaches up to $\sqrt{eB}\sim0.1$ GeV for RHIC and $\sqrt{eB}\sim 0.5$ GeV for the LHC~\cite{Skokov:2009qp}. This magnetic field is external since it is produced by the spectators, and though it has a very short lifetime (of the order of $1$ fm/$c$), the magnetic `impulse' coincides with the generation of the quark-gluon plasma and thus may have a significant effect on the properties of the transition.

For noncentral heavy ion collisions, an exciting consequence of the interplay between the strong magnetic field and the nontrivial topological structure of the quark-gluon plasma is the so-called chiral magnetic effect~\cite{Kharzeev:2007jp,Fukushima:2008xe}. This effect creates an electric current of quarks (anti)parallel to the external magnetic field, which may result in a preferential emission of charged particles perpendicular to the reaction plane, leading to event-by-event CP-violation~\cite{Kharzeev:2004ey}. Recent measurements from the STAR experiment at RHIC~\cite{Voloshin:2008jx,:2009uh} and the ALICE experiment at the LHC~\cite{Selyuzhenkov:2011xq} are in qualitative agreement with this picture, however, the interpretation of these results is still under discussion~\cite{Wang:2009kd,Muller:2010jd,Voronyuk:2011jd}.

Because of this high phenomenological relevance, the effect of a finite magnetic field on the strong interactions has been studied extensively in the last years, both using model calculations and lattice simulations. 
In particular, the structure of the QCD phase diagram in the $B-T$ plane has received increasing attention recently.
Calculations have been carried out within various low energy effective models of QCD. In the linear sigma model coupled to quarks and the Polyakov loop it was observed that the transition temperature increases with $B$~\cite{Mizher:2010zb}. Furthermore, a splitting between the deconfinement and chiral transitions was predicted to take place for large external fields. The strength of the transition was also observed to increase, which eventually results in a first-order phase transition~\cite{Fraga:2008um}. Similar conclusions with respect to the increase in $T_c$ and in the strength of the transitions were drawn from studies of the Nambu-Jona-Lasinio model and extended versions thereof, like the EPNJL and PNJL$_8$ models~\cite{Gatto:2010pt,Gatto:2010qs}, see also~\cite{Osipov:2007je}, and the nonlocal PNJL model~\cite{Kashiwa:2011js}. 

The presence of the external magnetic field was also shown to increase the transition temperature within other types of models like in the Sakai-Sugimoto model of large $N_c$ gauge theories~\cite{Johnson:2008vna}, in the Gross-Neveu model in lower dimensions~\cite{Kanemura:1997vi,Klimenko:1992ch}, in 2+1 dimensional QED when described by Schwinger-Dyson equations~\cite{Alexandre:2000yf} and within the holographic approach~\cite{Evans:2010xs}. However, the opposite effect of a decreasing deconfinement transition temperature was predicted using chiral perturbation theory for two quark flavors~\cite{Agasian:2008tb}. 
A decrease in $T_c$ was also observed in the linear sigma model if the quark vacuum contributions are neglected~\cite{Mizher:2010zb} and in the Sakai-Sugimoto model with nonzero chemical potential~\cite{Preis:2010cq}.
We mention that lattice simulations indicate a reduction of
the transition temperature of QCD in an external \emph{chromo}magnetic
field~\cite{Cea:2005td,Cea:2002wx,Cea:2007yv}.

The phase diagram is in most cases predicted by studying chiral symmetry breaking, i.e. the behavior of the chiral condensate or the dynamical quark mass as a function of $B$. Most of the low energy models agree that chiral symmetry breaking is enhanced as the magnetic field $B$ grows~\cite{Gusynin:1995nb,Nam:2011vn,Boomsma:2009yk}; in particular the value of the chiral condensate was found to increase linearly with $|B|$ in leading order~\cite{Shushpanov:1997sf,Cohen:2007bt,Agasian:2001hv}. The condensate also increases with $B$ -- although with a quadratic leading order -- within the AdS/CFT duality picture~\cite{Zayakin:2008cy}, and with $B^{3/2}$ in holography~\cite{Evans:2010xs}. On the other hand, it was also conjectured that the running of the strong coupling in the presence of magnetic fields may modify this magnetic catalysis, and even turn the effect around to make the dynamical mass decrease with $B$ in some regions~\cite{Miransky:2002rp}.

In recent lattice simulations with $N_f=2$ flavors of staggered quarks~\cite{D'Elia:2010nq} the chiral condensate was observed to grow with the external field for any temperature $T$ in the transition region. The size of this effect was however found to be different for different values of $T$, resulting in an increase in both the pseudocritical temperature $T_c$ and the strength of the transition. Furthermore, according to the findings of~\cite{D'Elia:2010nq}, the relative change in $T_c$ is of the order of one percent for several larger-than-physical pion masses.

In this paper our aim is to perform a similar lattice study, but with improved gauge and smeared fermionic actions and with $N_f=2+1$ flavors of quarks, at the physical pion mass, and extrapolate the results to the continuum limit. We include the magnetic field in the fermion determinant and study its effect dynamically to investigate how the strength and the pseudocritical temperature of the QCD transition change as the external magnetic field is switched on. We explore a wide temperature region around the zero-field pseudocritical temperature $T_c(B=0)$, for various values of the magnetic field, ranging from $\sqrt{eB}\sim100$ MeV to $\sqrt{eB}\sim 1$ GeV, i.e. covering the regions that are phenomenologically interesting for noncentral heavy ion collisions and for the evolution of the early universe.

This paper is structured as follows: first the implementation of the magnetic field on the lattice is described. Then we define the observables of interest, including the chiral condensate, chiral susceptibility and strange quark number susceptibility, and discuss their renormalization at zero and nonzero $B$. After presenting the simulation setup and the details of the analysis we show our results for the transition temperature and the width of the transition.

\section{Magnetic field on the lattice}

Let us consider the case of a constant external magnetic field $\mathbf{B}=(0,0,B)$, that is pointing in the $z$ direction. In the continuum such a magnetic field can be realized by, e.g., the following vector potential,
\be
A_\nu = ( \mathbf{A}, A_t ) =  ( 0, B x, 0, 0 ).
\label{eq:contvecpot}
\ee
Any other vector potential satisfying $\mathbf{B}=\textmd{curl}(\mathbf{A})$ corresponds to the same physical system, and is connected to the above choice by an appropriate $\mathrm{U}(1)$ gauge transformation.

It is well known that in a finite box with periodic boundary conditions the magnetic flux cannot be arbitrary, but is quantized in terms of the area $A$ of the system in the plane orthogonal to the external field~\cite{'tHooft:1979uj,AlHashimi:2008hr}. This leads to the quantization condition,
\be
qB\cdot A = 2\pi N_b, \quad\quad\quad N_b\in\mathds{Z},
\label{eq:quantcont}
\ee
where $q$ is the charge of the particle. In turn, on the lattice the area $A$ is also quantized as $A=N_xN_y a^2$, where $a$ is the lattice spacing (we restrict the discussion to isotropic lattices) and $N_\nu$ is the number of lattice points in the direction $\nu$. This implies that the lattice discretization also imposes an upper bound on the magnetic flux. To see this explicitly, let us write down how the continuum vector potential~(\ref{eq:contvecpot}) can be represented by complex phases $u_\nu(n)\in \mathrm{U}(1)$ that multiply the $U_\nu(n)\in \mathrm{SU}(3)$ links of the lattice,
\be
%\begin{split}
%u_y(n) &= e^{i a^2 q B n_x}, \quad\quad\quad\quad\quad\quad u_x(N_x-1,n_y,n_z,n_t) = e^{-i a^2 q B N_x n_y}, \\
%u_x(n) &= 1, \quad\quad n_x\ne N_x-1, \quad\quad\quad\quad 
%u_\nu(n) =1, \quad\quad\quad\quad \nu\not\in\{x,y\},
%\end{split}
%\label{eq:links2}
%\ee
%\be
\begin{split}
u_y(n) &= e^{i a^2 q B n_x}, \\
u_x(N_x-1,n_y,n_z,n_t) &= e^{-i a^2 q B N_x n_y}, \\
u_x(n) &= 1, \quad\quad\quad\quad\quad\quad\quad n_x\ne N_x-1, \\
u_\nu(n) &=1, \quad\quad\quad\quad\quad\quad\quad \nu\not\in\{x,y\},
\end{split}
\label{eq:links2}
\ee
where the sites are labeled by integers $n=(n_x,n_y,n_z,n_t)$, with $n_\nu=0\ldots N_\nu-1$. 
Constant magnetic background fields were first used in lattice studies of nucleon magnetic moments at zero temperature~\cite{Martinelli:1982cb,Bernard:1982yu,Zhou:2002km}. At finite temperature this approach was first
realized in~\cite{D'Elia:2010nq, Roberts:2010cz}.

On the lattice the quantization condition~(\ref{eq:quantcont}) thus takes the same form as in the continuum with the area discretized in terms of the lattice spacing,
\be
qB\cdot a^2 = \frac{2\pi N_b}{N_x N_y}.
\label{eq:quantlatt}
\ee
In this formulation we have periodic boundary conditions in all spatial directions and the magnetic flux going through any plaquette in the $x-y$ plane is constant. Furthermore, this implementation of the magnetic field is periodic in the flux quantum $N_b$ with a period of $N_xN_y$, so its value is effectively constrained to $0\le N_b< N_xN_y$. This prescription is discussed in more detail in appendix~\ref{sec:app1}. We note that the periodicity of the field in $N_b$ implies a stronger constraint for the flux,
\be
0 \le N_b< \frac{N_xN_y}{4},
\ee
where the correspondence between the implementation and the actual value of $B$ is unambiguous. Namely, at larger values of $N_b$ the periodicity is expected to introduce saturation effects, like it was observed in~\cite{D'Elia:2011zu,Bruckmann:2011zx}. The largest possible magnetic field is therefore $qB^{\rm max} = \pi/2 \cdot a^{-2}$. 

On the lattice the temperature of the system is given by the inverse temporal extension as $T=(N_ta)^{-1}$. It is therefore clear from equation~(\ref{eq:quantlatt}) that the minimal value of the magnetic field is $qB^{\rm min} = T^2 \cdot 2\pi N_t^2/N_xN_y$. Thus, to increase the maximal field one has to decrease $a$, and to decrease the minimal magnetic field one has to increase $N_xN_y$.
Taking these considerations into account, with reasonable lattice spacings and lattice extensions, the lattice magnetic field covers the region $\sqrt{qB}=0.1 \ldots 2$ GeV.

We remark that if there are particles with different charges in the system, then the quantization condition for $B$ has to be fulfilled for the greatest common divisor -- in our case this is the down quark charge, $q=q_d=-|e|/3$ in equation~(\ref{eq:quantlatt}) with $e$ the charge of the electron. This will then determine the minimal field. Fortunately in nature the ratio of quark charges is a small natural number so that the up and down quarks can be studied together.
We note furthermore that the above implementation of the magnetic field leads to no sign problem, in contrast to a finite chemical potential, or a (Minkowskian) electric field. 

\section{Observables at finite \boldmath \texorpdfstring{$B$}{B}}
\label{sec:magnobs}

Let us consider the staggered partition function with three flavors $u,d$ and $s$. Each quark flavor has to be treated separately since the charges/masses are different: $q_u=-2q_d=-2q_s$, and we assume $m_u=m_d\ne m_s$. The partition function reads, after taking the fourth roots of the fermion determinants,
\be
\Z = \int \D U e^{-\beta S_g} \left[ \det M(U,q_u,m_u,\mu_u)\right]^{1/4} \left[\det M(U,q_d,m_d,\mu_d)\right]^{1/4} \left[\det M(U,q_s,m_s,\mu_s)\right]^{1/4},
\label{eq:partfunc}
\ee
where the fermion matrix is $M(U,q,m,\mu) = \slashed D(U,q,\mu) + m\mathds{1}$. (Here we do not address problems arising from the rooting trick~\cite{Durr:2005ax}). The dependence on the chemical potential is only made explicit to define derivatives of the partition function with respect to $\mu$, see equation~(\ref{eq:defc2s}), and later we set all chemical potentials to zero.
Since we are only concerned with a constant external field, the dynamics of the $\mathrm{U}(1)$ field introduced above does not have to be taken into account; in the gauge sector we only have the $\mathrm{SU}(3)$ kinetic term $S_g$ with inverse gauge coupling $\beta=6/g^2$	.

To study thermodynamics in a nonzero external field we analyze the chiral condensates and chiral susceptibilities for the light flavors $f=u,d$,
\be
\bar{\psi}\psi_f \equiv \frac{T}{V}\frac{\partial \log \Z}{\partial m_f}, \quad\quad\quad\quad\quad\quad \chi_{f} \equiv \frac{T}{V}\frac{\partial^2 \log \Z}{\partial m_f^2},
\ee
and the strange quark number susceptibility,
\be
c_2^s \equiv \frac{T}{V}\frac{1}{T^2}\frac{\partial^2 \log \Z}{\partial \mu_s^2},
\label{eq:defc2s}
\ee
where we defined the spatial volume of the system as $V= (N_sa)^3$ with $N_s\equiv N_x=N_y=N_z$.
The condensate for a particular flavor will be denoted in the following by the first letter of the flavor name, e.g. $\bar u u$. 

To take the continuum limit, the renormalization of these observables has to be carried out.
The logarithm of the partition function $\log\Z$ (i.e. the free energy) at $B=0$ contains additive divergences of the forms $a^{-4}$, $m^2a^{-2}$ and $m^4\log(a)$~\cite{Leutwyler:1992yt}.
In section~\ref{sec:renorm} we will show -- based on the behavior of the beta function measured at zero temperature -- that there are no additional $B$-dependent divergences.
Therefore the additive divergences of the observables derived from the free energy can be eliminated by subtracting the $T=0$, $B=0$ contribution. 
In the chiral quantities there are also multiplicative divergences caused by the derivative with respect to the quark mass.
To eliminate this multiplicative divergence in the chiral condensate (susceptibility), we multiply by the first (second) power of the bare quark mass~\cite{Endrodi:2011gv}. Finally, to obtain a dimensionless combination we divide by the fourth power of the $T=0$ pion mass $m_\pi^4$,
\be
\begin{split}
\bar{\psi}\psi_f^r (B,T) &= m_f \left [\bar{\psi}\psi_f(B,T) - \bar{\psi}\psi_f(B=0,T=0) \right] \frac{1}{m_\pi^4}, \\
\chi_f^r(B,T) &= m_f^2 \big [\chi_f(B,T) - \chi_f(B=0,T=0) \big] \frac{1}{m_\pi^4}.
\end{split}
\ee
Note that this procedure leads to a renormalized condensate that, for $B=0$, is zero at $T=0$ and approaches a negative value as $T$ is increased.

Considering the strange quark number susceptibility, $c_2^s$ needs no renormalization (neither at $B=0$ nor at $B\ne 0$) since it is connected to a conserved current.

\section{Renormalization at finite \boldmath \texorpdfstring{$B$}{B}}
\label{sec:renorm}

We expect that a nonzero external magnetic field does not introduce new divergences in the free energy density, since the external field -- just like a chemical potential -- is coupled to the current $\bar\psi\gamma_\nu \psi$ which is conserved. This expectation is also supported by the fact that in the presence of the external field there are no additional, divergent Feynman-diagrams due to the absence of internal photon lines.\footnote{Consider e.g. the gluon self-energy diagram (with one quark loop) which is -- in a gauge invariant regularization -- logarithmically divergent. The coupling to the external magnetic field in the lowest order in $B$ is given by two external photon legs attached to the quark loop. This diagram is clearly finite since it contains two extra quark propagators.}
Moreover, the vacuum energy was also calculated in the effective potential approach~\cite{Salam1975203} and its divergent part was found to be independent of $B$.
The absence (or presence) of $B$-dependent divergences in the free energy density is closely related to the non-renormalization (or renormalization) of $B$ itself. 
In fact, in a gauge invariant renormalization scheme the product $eA_\mu$ needs no renormalization because of the $\mathrm{U}(1)$ Ward-Takahashi identity (see appendix~\ref{app:WT}).
In our case the magnetic field always appears in the combination $eB$ and, therefore, we expect that it is not subject to renormalization\footnote{Note that for a dynamical $\mathrm{U}(1)$ theory, $B$ would appear separately in the photon gauge action. This is, however, not the case for the present study.}.
We now check these expectations numerically.

From the point of view of renormalization theory it may be instructive to draw a parallel between the magnetic field and the quark mass. {\it If} the magnetic field were to induce new divergences (like the mass does) then $eB$ itself (like $m$) would be subject to renormalization as $(eB)^r=Z\cdot (eB)$ with $Z$ a corresponding renormalization constant. 
However, in the lattice approach the magnetic field -- unlike the mass -- fulfills a quantization condition, as in equation~(\ref{eq:quantlatt}). Therefore the renormalization of $eB$ can only amount to a shift in the lattice spacing $a$, such that $a^2eB$ = $a_{\rm shifted}^2 (eB)^r$ is satisfied.
This implies that the lattice scale has to change if the renormalization of the magnetic field is nontrivial, i.e. if there are $eB$-dependent divergences. This is expected since the lattice scale is determined by the beta-function of the theory which is given in terms of the renormalization scale-dependent, divergent Feynman-diagrams (see e.g.~\cite{Peskin:1995ev}). For the magnetic field however, due to the quantization condition, the {\it only} possible effect of such divergent diagrams is to alter the lattice scale.

Therefore we propose to measure a physical quantity $\phi$ at $T=0$ as a function of the magnetic flux $N_b$ for different lattice spacings. We take the lattice scale $a(\beta)$ and the line of constant physics (LCP) $m(\beta)$ which are measured at $N_b=0$ (see section~\ref{sec:simdet}), and assume that they are also valid at $N_b>0$. We use the scale to determine the magnetic field from the flux according to equation~(\ref{eq:quantlatt}) and the quantity $\phi$ in physical units. Then we compare the running of $\phi(eB,a)$ with the magnetic field for different lattice spacings. If this quantity has a meaningful continuum limit,
\be
\lim_{a\to 0} \phi(eB,a) = \phi^{\rm cont}(eB),
\ee
i.e. if for small enough lattice spacings the dependence of $\phi(eB,a)$ on $a$ is suppressed, then our assumption was valid and the lattice scale $a(\beta)$ and the LCP $m(\beta)$ are also correct for $N_b>0$. In view of the discussion in the previous paragraph, this suggests that there are no $eB$-dependent divergences. In the opposite case the lattice scale does depend on $N_b$, which in turn would indicate the presence of $eB$-dependent divergences.

For physical quantities we choose the charged pion mass $\phi=m_{\pi^+}$ and the Sommer parameter $\phi=r_0$. For their definition and measurement details see, e.g.~\cite{Aoki:2009sc}. For the charged pion mass we expect a strong dependence on the magnetic field, in the form~\cite{Chernodub:2010qx},
\be
m_{\pi^+}(B) = \sqrt{m_{\pi^+}^2(0) + |eB|},
\label{eq:chargedpion}
\ee
which can be deduced from the dispersion relation for spin-1 mesons\footnote{We note that this expression ought to receive corrections for large $B$ due to pair production.}.
On the other hand, for the Sommer parameter, which is defined using the potential between static color charges, we expect the dependence on $eB$ to be suppressed.

\begin{figure}[ht!]
\centering
\vspace*{-0.2cm}
\includegraphics*[width=7.4cm]{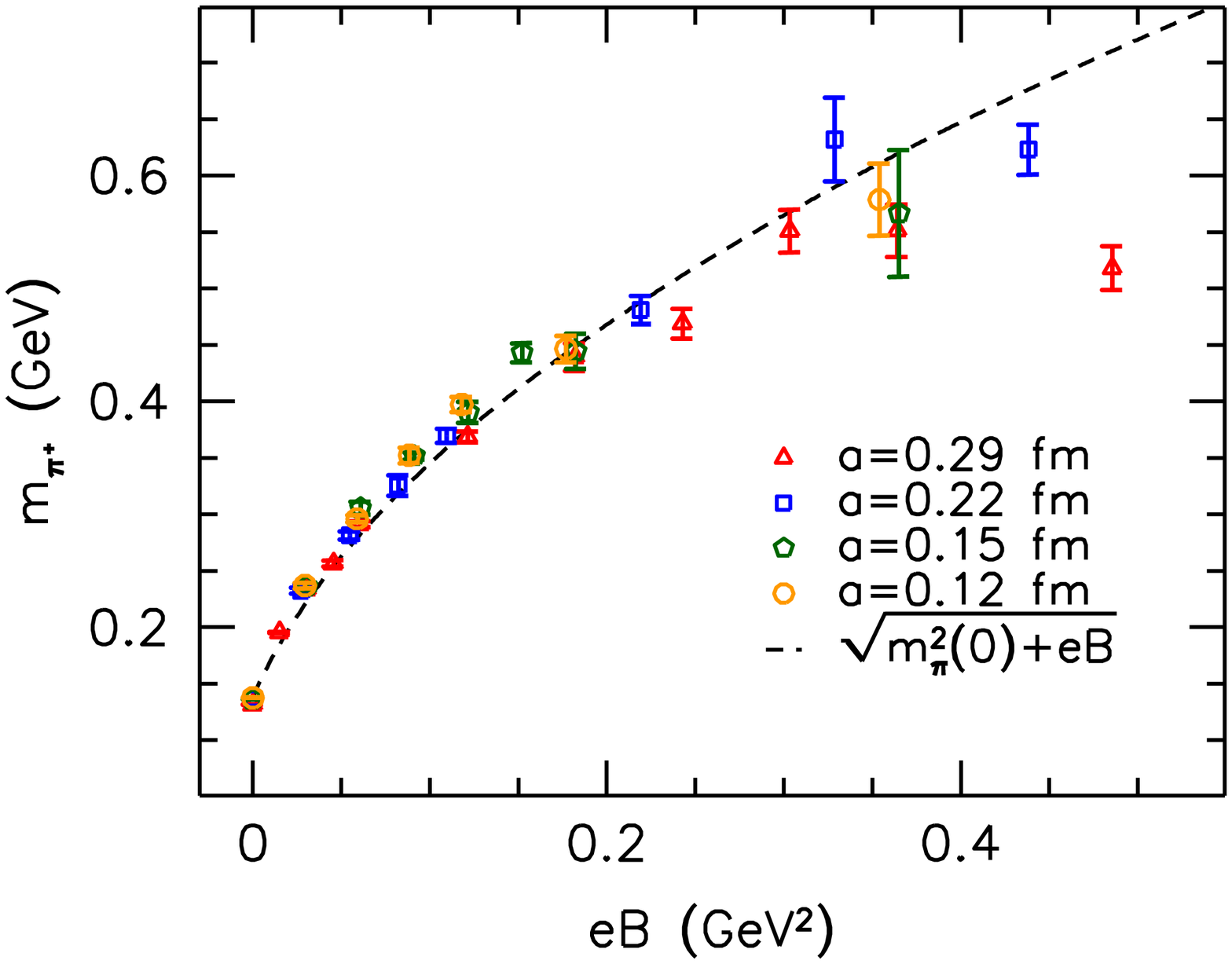}
\includegraphics*[width=7.4cm]{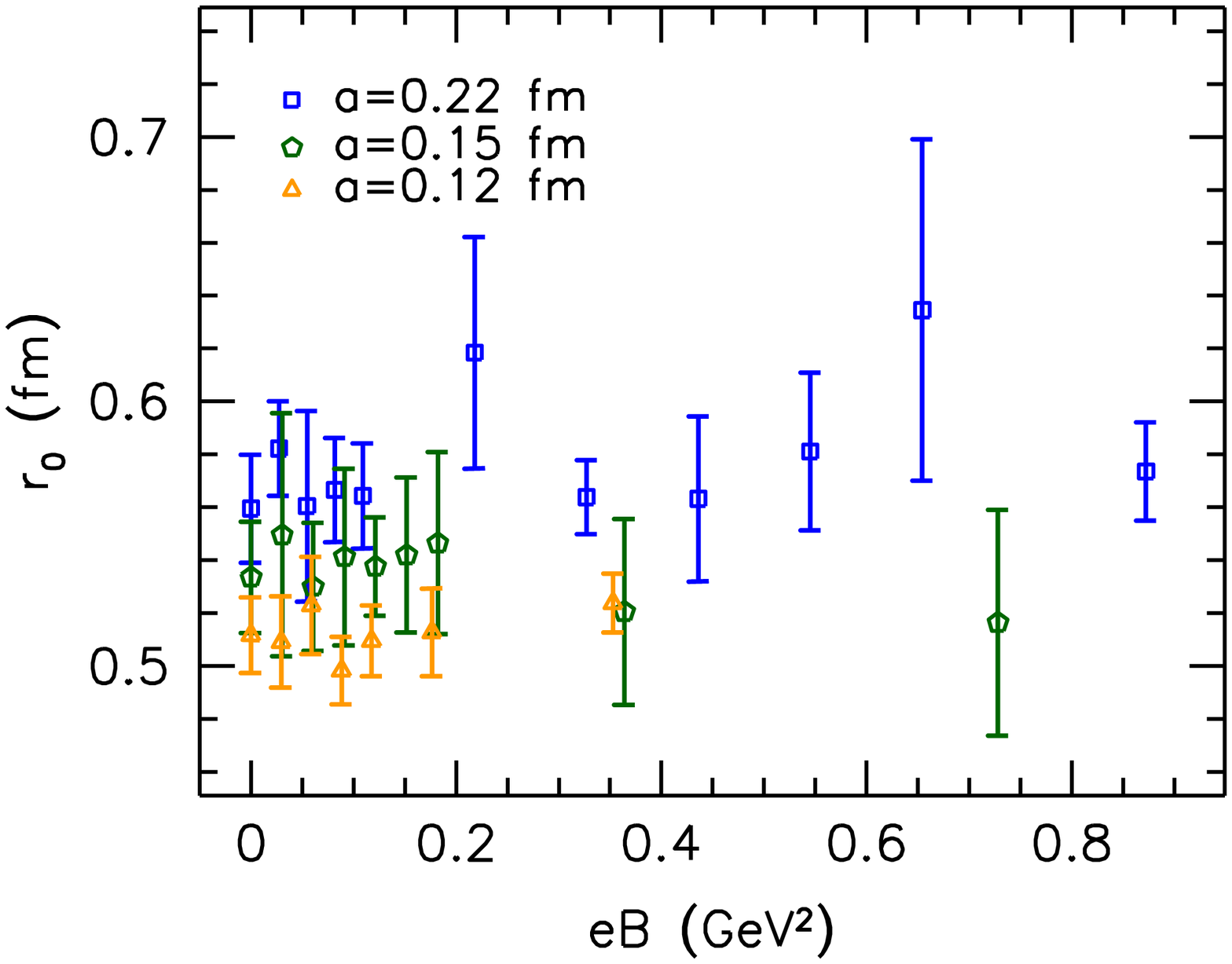}
\vspace*{-0.1cm}
\caption{The mass of the charged pion (left panel) and the Sommer parameter (right panel) as functions of the external magnetic field for different lattice spacings. Results for the pion mass are compared to the analytic prediction (see text). The good scaling of the lattice results and the independence of $r_0$ on $eB$ indicate the absence of $eB$-dependent divergences.}
\label{fig:zeroT}
\end{figure}

In figure~\ref{fig:zeroT} the mass of the charged pion and the Sommer parameter are plotted as functions of the external field for various lattice spacings, at $a=0.29$, $0.22$, $0.15$ and $0.12$ fm (for the Sommer parameter the coarsest lattice spacing is not shown since here $r_0$ has large systematic errors). The lattice geometries and simulation parameters for these runs are tabulated in appendix~\ref{sec:simpar}. At $B=0$ we used the measurements presented in~\cite{Aoki:2009sc}.
We observe a nice scaling with $a$ for both quantities, in the region $eB\lesssim 0.4 \textmd{ GeV}^2$, where we have data for the three finest lattices. Results are consistent with a constant behavior for the Sommer parameter which indicates that the lattice spacing is not modified by the external field beyond our statistical accuracy.

Data for the charged pion mass are also as expected and agree with the analytic prediction~(\ref{eq:chargedpion}) within $2-3\%$. For large $N_b\sim a^2eB$ we see deviations from the continuum scaling only for the coarsest lattice which is most probably due to lattice artefacts stemming from the periodicity of the lattice magnetic field~(\ref{eq:links2}). Based on theoretical arguments (see appendix~\ref{app:WT}) and on these observations we conclude that it is safe to use the $B=0$ lattice scale and LCP at nonzero external fields, and to exclude the possibility of $eB$-dependent divergences.

\section{Simulation details}
\label{sec:simdet}

We use the tree-level improved Symanzik gauge action and stout smeared staggered fermions; details about the action can be found in~\cite{Aoki:2005vt}. We generate lattice configurations both at $T=0$ and $T>0$ with an exact RHMC algorithm, for various values of the gauge coupling and the magnetic flux.
(To discretize the external magnetic field, the smeared links are multiplied by the $\mathrm{U}(1)$ links of equation~(\ref{eq:links2}).)
For our zero temperature measurements we simulate $24^3\times 32$, $32^3\times 48$ and $40^3\times 48$ lattices, while for the finite temperature runs we have lattice configurations with $N_t=6,8$ and $10$. Finite volume effects are studied on the $N_t=6$ ensemble using sets of $N_s=16$, $24$ and $32$ lattices. The masses of the up, down and strange quarks are set to their physical values along the line of constant physics (LCP) by fixing the ratios $f_K/m_\pi$ and $f_K/m_K$ to their experimental values. The lattice spacing is determined by $f_K$. Details of the determination of the LCP and the lattice scale can be found in, e.g.~\cite{Borsanyi:2010cj}. Based on the reasoning presented in section~\ref{sec:renorm}, we use the lattice spacing measurements at $T=0$ and $B=0$ to set the scale also at $T \ne 0$ and $B\ne 0$. The nonzero value of the magnetic field may of course modify e.g. the pion decay constant, just as the temperature can, but this is not important from the aspect of matching the lattice quantities at $T=0$, $B=0$ to their experimental values (which are also measured at $T=0$ and $B=0$).

Altogether we generated several hundred to few thousand thermalized trajectories for each $\beta$ and $N_b$ (see list of simulation parameters in appendix~\ref{sec:simpar}), and performed the measurements on every fifth one to decrease autocorrelations.
The observables presented in section~\ref{sec:magnobs} were measured using the random estimator method, with 40 random vectors.
The production of configurations and the measurements were performed on CUDA-capable GPU clusters at the E\"otv\"os University in Budapest and on the Bluegene/P at FZ J\"ulich.

We mention here that the staggered formulation of fermions introduces lattice artefacts due to the splitting of hadron states into multiplets with different masses~\cite{Ishizuka:1993mt}. We keep the lowest lying pion state at the physical pion mass, while the other members of the pion multiplet are heavier. In the continuum limit this mass splitting between the tastes vanishes. However, at finite lattice spacing it can distort thermodynamic quantities. To reduce this splitting we apply stout smearing in the fermionic action, which is known to significantly reduce taste symmetry violation~\cite{Borsanyi:2010cj}.

\section{Analysis details}
\label{sec:analysis}

To study the $B$-dependence of the observables of section~\ref{sec:magnobs}
we scan a wide interval in both the temperature $T$ and the flux quantum $N_b$. The latter is proportional to $eB/T^2$, see equation~(\ref{eq:quantlatt}), so, since the transition spreads over a wide temperature region, the physical magnetic field also changes by up to a factor of two along an $N_b=\textmd{const.}$ line between $T=120$ MeV and $T=180$ MeV. To correct for this change one can simulate at parameters $T,N_b$ tuned such that the physical magnetic field remains the same. However, since $N_b$ cannot be varied continuously, here we follow a slightly different approach.

\begin{wrapfigure}{r}{7.0cm}
\centering
\vspace*{-0.5cm}
\includegraphics*[width=7.0cm]{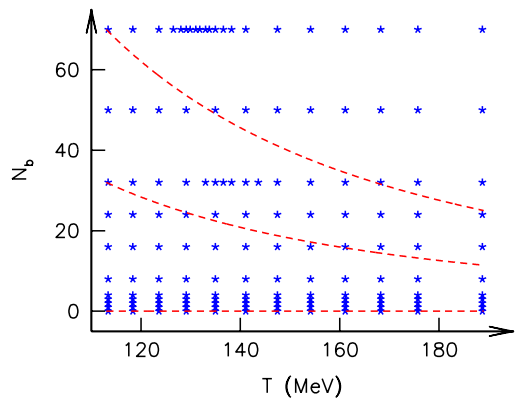}
\vspace*{-0.8cm}
\caption{Our simulation points on $24^3\times 6$ lattices (blue crosses) and the lines of constant magnetic field (red dashed lines).}
\label{fig:plane}
\vspace*{-0.2cm}
\end{wrapfigure}

\begin{figure}[h!]
\centering
\mbox{
\includegraphics*[height=5.8cm]{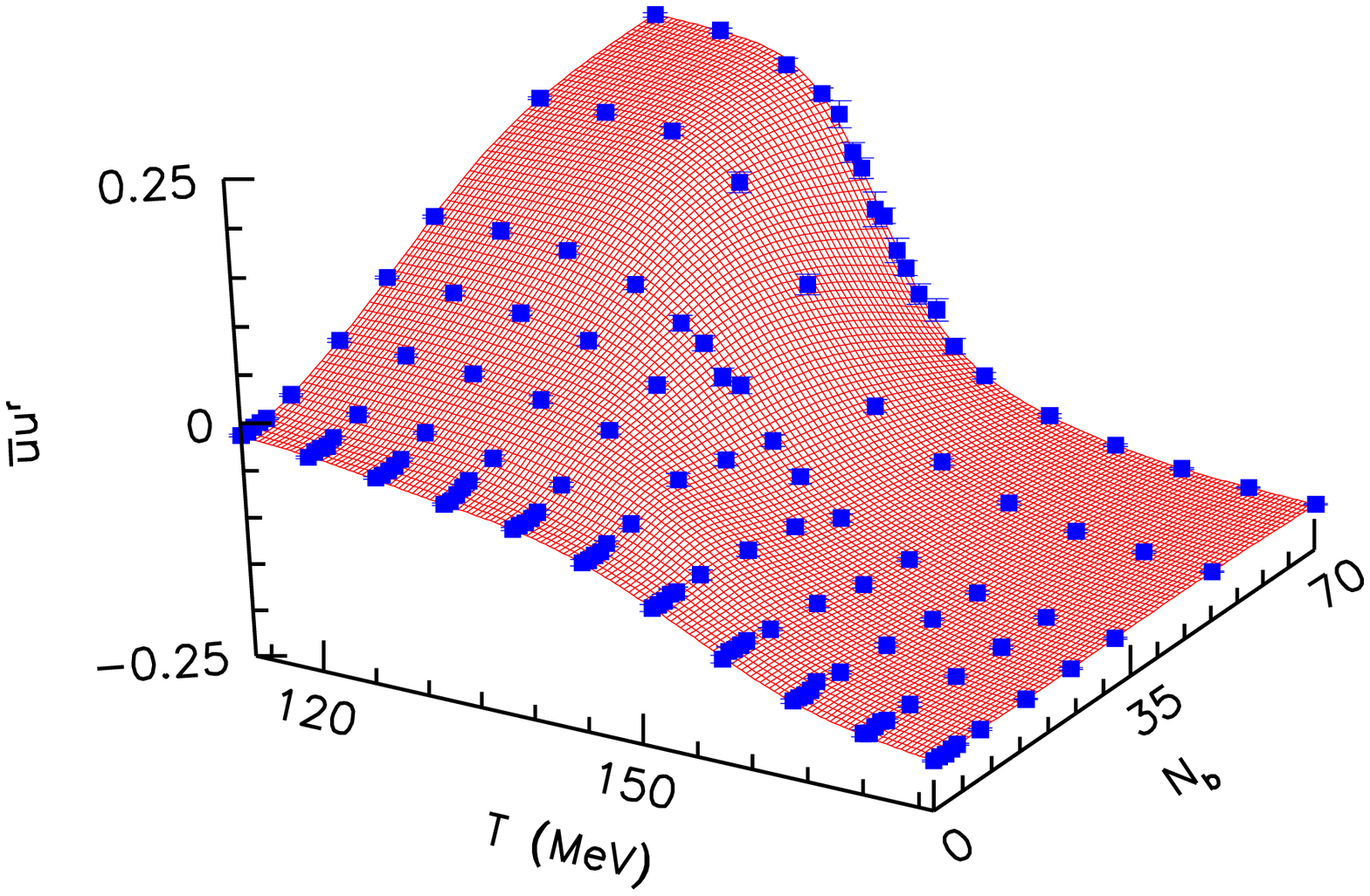}
\includegraphics*[height=5.8cm]{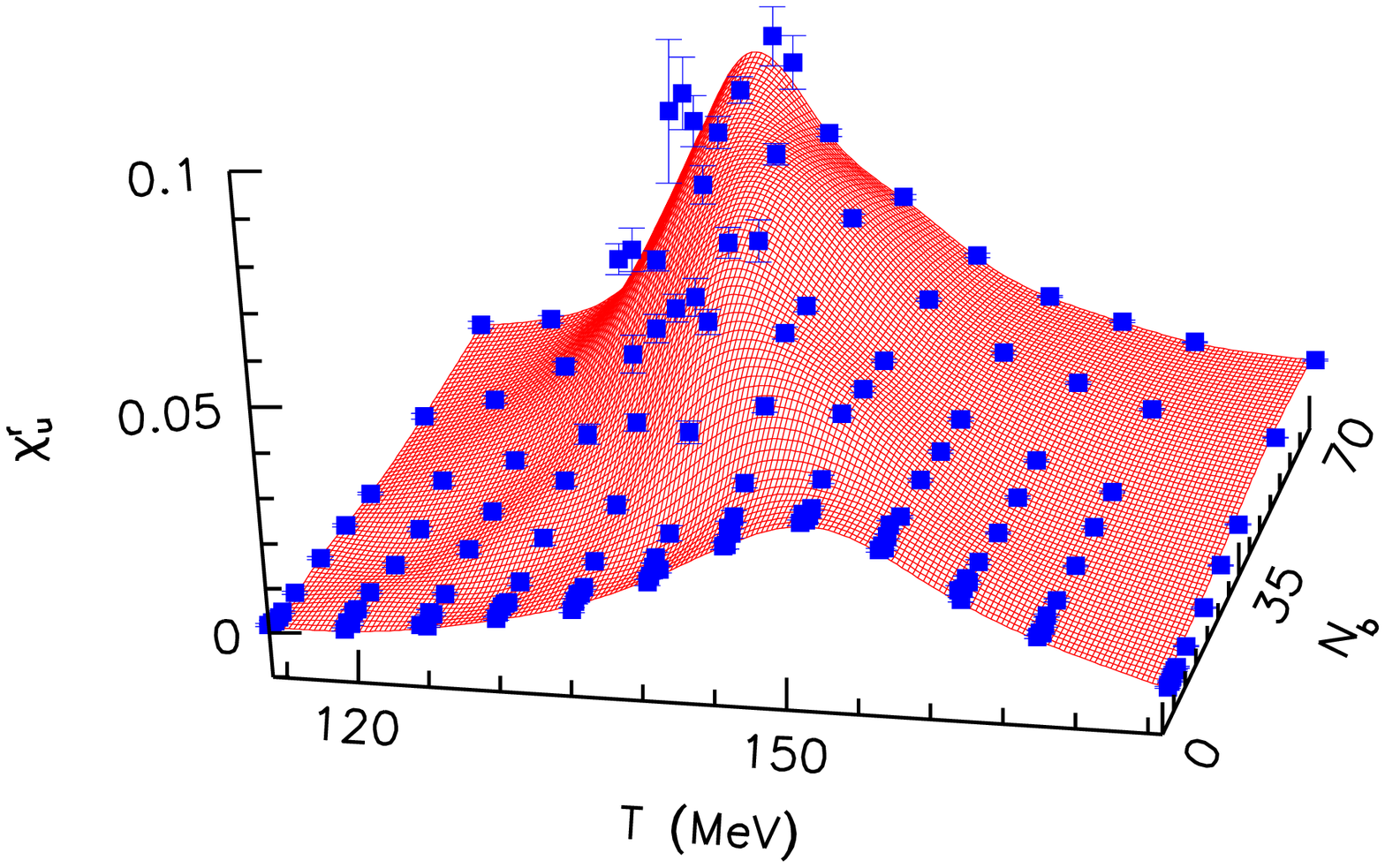}
}
\mbox{
\includegraphics*[height=5.8cm]{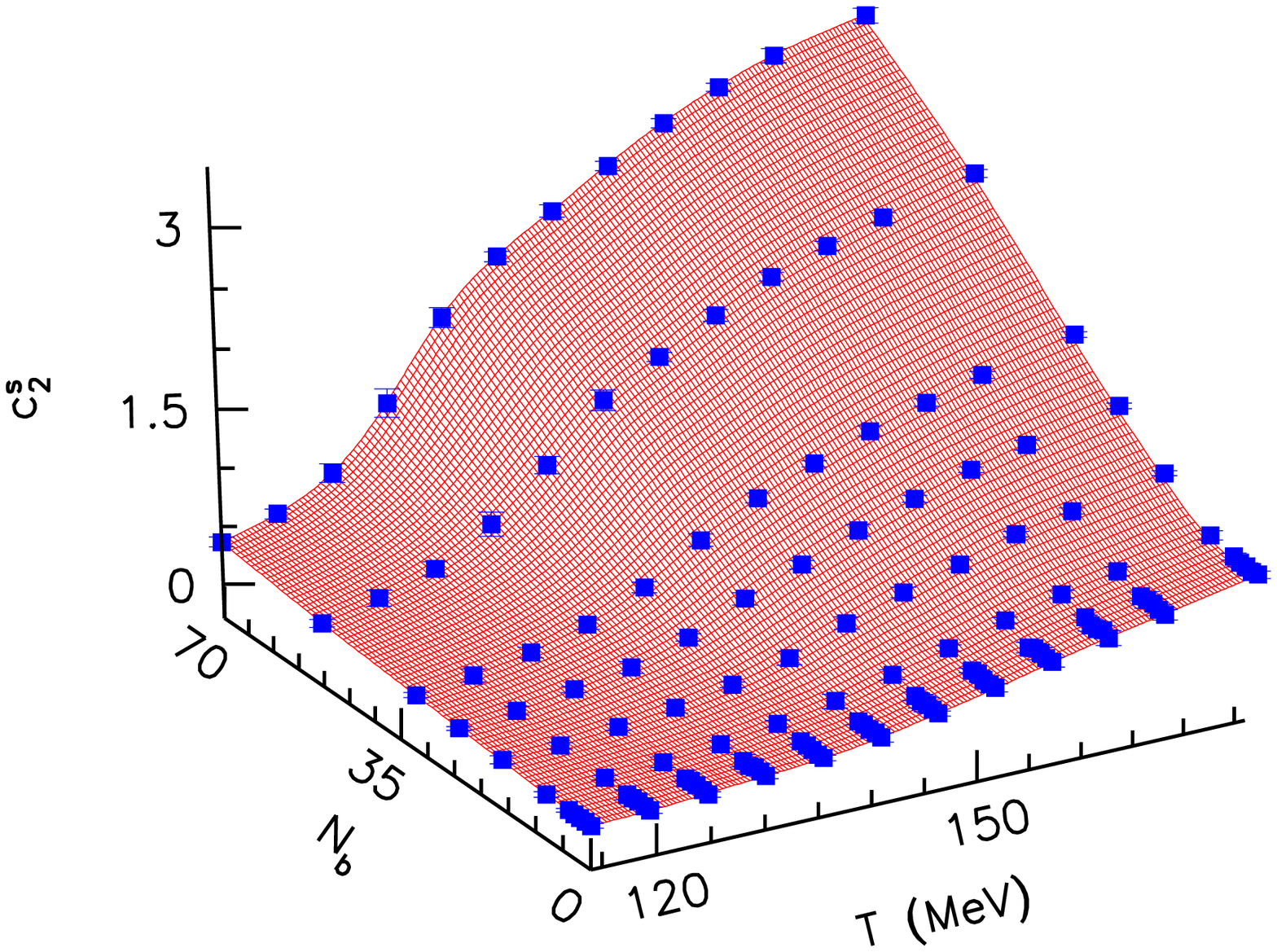}
}
\caption{The renormalized up quark condensate (upper left panel), its susceptibility (upper right panel),
and the strange susceptibility (lower panel) as functions of $T$ and $N_b$ on our $N_t=6$ lattices (note that viewpoints are different in order to better show the interesting structures in the particular observables). Measurements are denoted by the blue points, while the red surface is the spline fit to the data. The corresponding fit qualities are $\chi^2/\textmd{dof.} \approx 1.8$, 1.5 and 1.2, respectively.}
\label{fig:surface}
\end{figure}

We measure our observables along a grid of points in the $T-N_b$ plane, as depicted in figure~\ref{fig:plane}. The simulation points are denoted by the blue crosses, while the $eB=\textmd{const.}$ curves are shown by the red dashed lines. To perform the interpolation of the measurements along these lines in a systematic and effective way, we fit a two-dimensional spline function to the data points. A similar approach is described in~\cite{Endrodi:2010ai} for the fitting of the gradient of a two-dimensional function. In figure~\ref{fig:surface} we show the observables as functions of $T$ and $N_b$ for our $N_t=6$ lattices. We obtain reliable results with good fit qualities; $\chi^2/\textmd{dof.}$ being in the range $1.2-1.8$.

We perform simulations over the same physical temperature and magnetic field range for two smaller lattice spacings at $N_t=8$ and $N_t=10$, with very similar $\chi^2/{\rm dof.}$ values for the spline fits as above. We use these three lattice spacings (around $T_c(0)$ they correspond roughly to $a=0.2$, 0.15 and 0.12 fm) to extrapolate our results to the continuum limit.

\section{Behavior of the condensate}
\label{sec:condensate}

We remark already at this point that the pseudocritical temperature -- as probably best visible in the upper right panel of figure~\ref{fig:surface} for the chiral susceptibility -- apparently decreases with increasing $N_b\sim B$, thereby contradicting a vast number of model calculations, see the summary given in the introduction. Furthermore this observation also disagrees with the lattice result of~\cite{D'Elia:2010nq}. First of all, to check our simulation code we reproduced the results of~\cite{D'Elia:2010nq} at a couple of points, see appendix~\ref{sec:codecheck}. Since we find a perfect agreement, we conclude that we are left with three possible reasons for the discrepancy. First, the lattice spacing of~\cite{D'Elia:2010nq} is larger, $a\approx 0.3$ fm, and also an unimproved action is used, so lattice discretization errors may be significant. Second, the present study uses $N_f=2+1$ flavors as opposed to the $N_f=2$ of~\cite{D'Elia:2010nq}, and the pseudocritical temperature is known to depend on the number of flavors~\cite{Karsch:2000kv}, which may also introduce systematic differences in the dependence on the external field. Third, the quark masses of~\cite{D'Elia:2010nq} are larger than in the present study, which can also cause drastic changes in thermodynamics -- for example the nature of the transition at $B=0$ depends very strongly (and non-monotonically) on the quark masses.

\begin{figure}[ht!]
\centering
\includegraphics*[width=7.4cm]{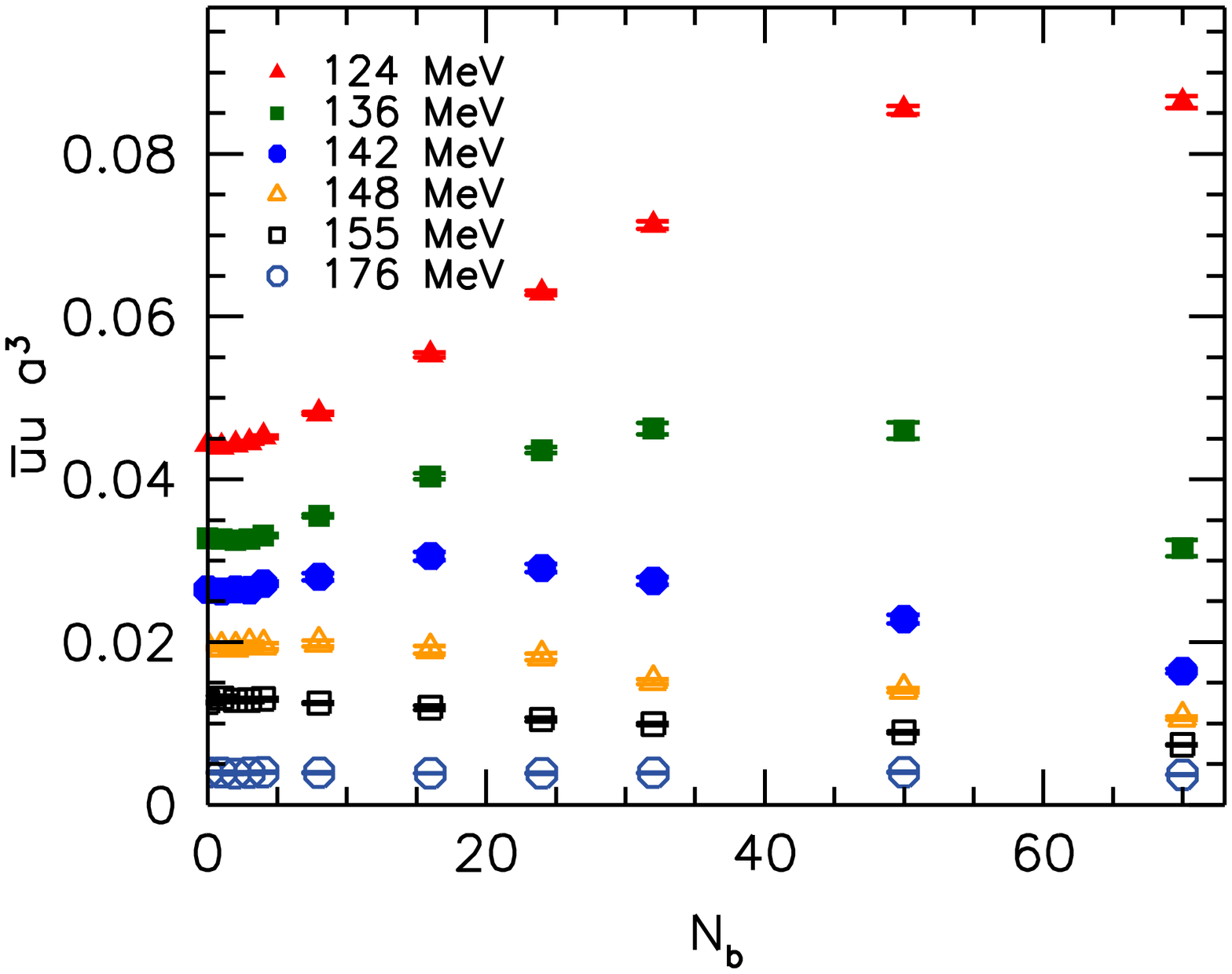}\quad
\includegraphics*[width=7.4cm]{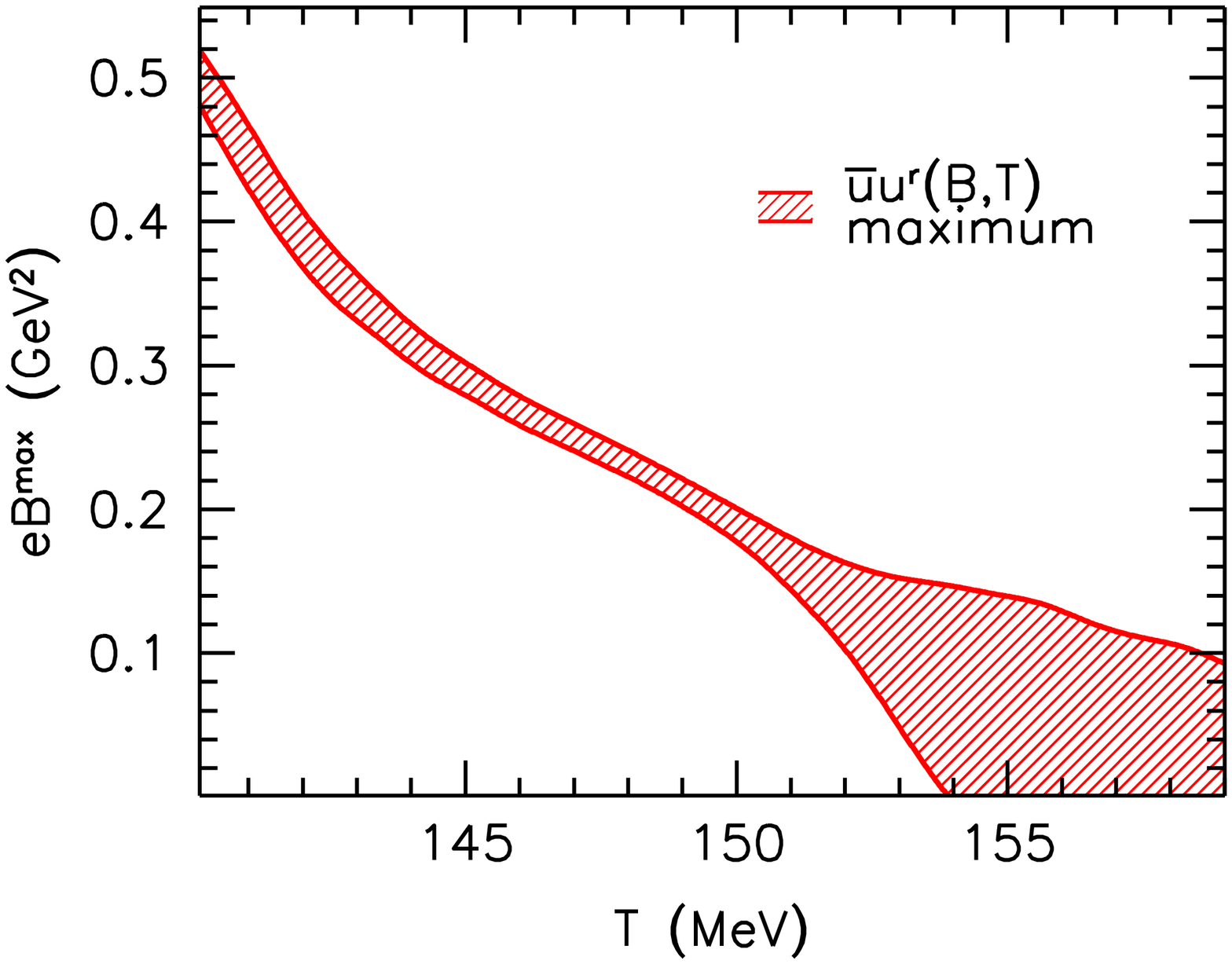}
\caption{The unrenormalized chiral condensate as a function of the flux quantum for various temperatures around the transition for $N_t=6$. A complex dependence $\bar u u(T,N_b)$ is observed, since in the deconfined phase in some regions the condensate decreases with growing $N_b$ (left panel). The magnetic field $B^{\rm max}$ where the renormalized condensate $\bar u u^r (T,B) $ is maximal, as a function of the temperature (right panel), as measured on $N_t=6$ lattices.}
\label{fig:pbp_beta}
\end{figure}

On closer inspection, the differences between our results and those of~\cite{D'Elia:2010nq} can actually be traced back to the behavior of the chiral condensate as a function of $B$ for a given temperature. While the authors of~\cite{D'Elia:2010nq} observed that at any temperature the condensate increases with $B$, we find that this dependence is more complex, see the left panel of figure~\ref{fig:pbp_beta} for our $N_t=6$ results. At $T=155$ MeV, which is just above the zero-field pseudocritical temperature, the bare condensate decreases by a factor of 2 between $N_b=0$ and $N_b=70$. As the temperature is reduced the $\bar u u(N_b)$ function starts to develop a maximum, clearly visible for $T=142$ MeV and $T=136$ MeV. This non-monotonic behavior is not due to the saturation effects caused by the periodic implementation of the magnetic field on the lattice, since this maximum is located at very different values of $N_b$ for temperatures differing only by a few percent. Furthermore, for high temperatures the decrease is already visible at $N_b<10$ which is in the first 5 percent of the period, even for the up quark. To better illustrate this effect and to show that renormalization and conversion from $N_b$ to $B$ does not change the picture qualitatively, in the right panel of figure~\ref{fig:pbp_beta} we plot the value of the external field $B^{\rm max}$ where the renormalized chiral condensate takes its maximum, as a function of the temperature. At high temperatures this maximum is located at $B^{\rm max}=0$, while below $T=155$ MeV it shifts to a nonzero magnetic field, in accordance with the left panel of figure~\ref{fig:pbp_beta}. As already mentioned in the introduction, the possibility of such a decrease in the condensate with $B$ was also raised in low energy model calculations~\cite{Miransky:2002rp}.

\begin{figure}[h!]
\centering
\mbox{
\includegraphics*[width=7.0cm]{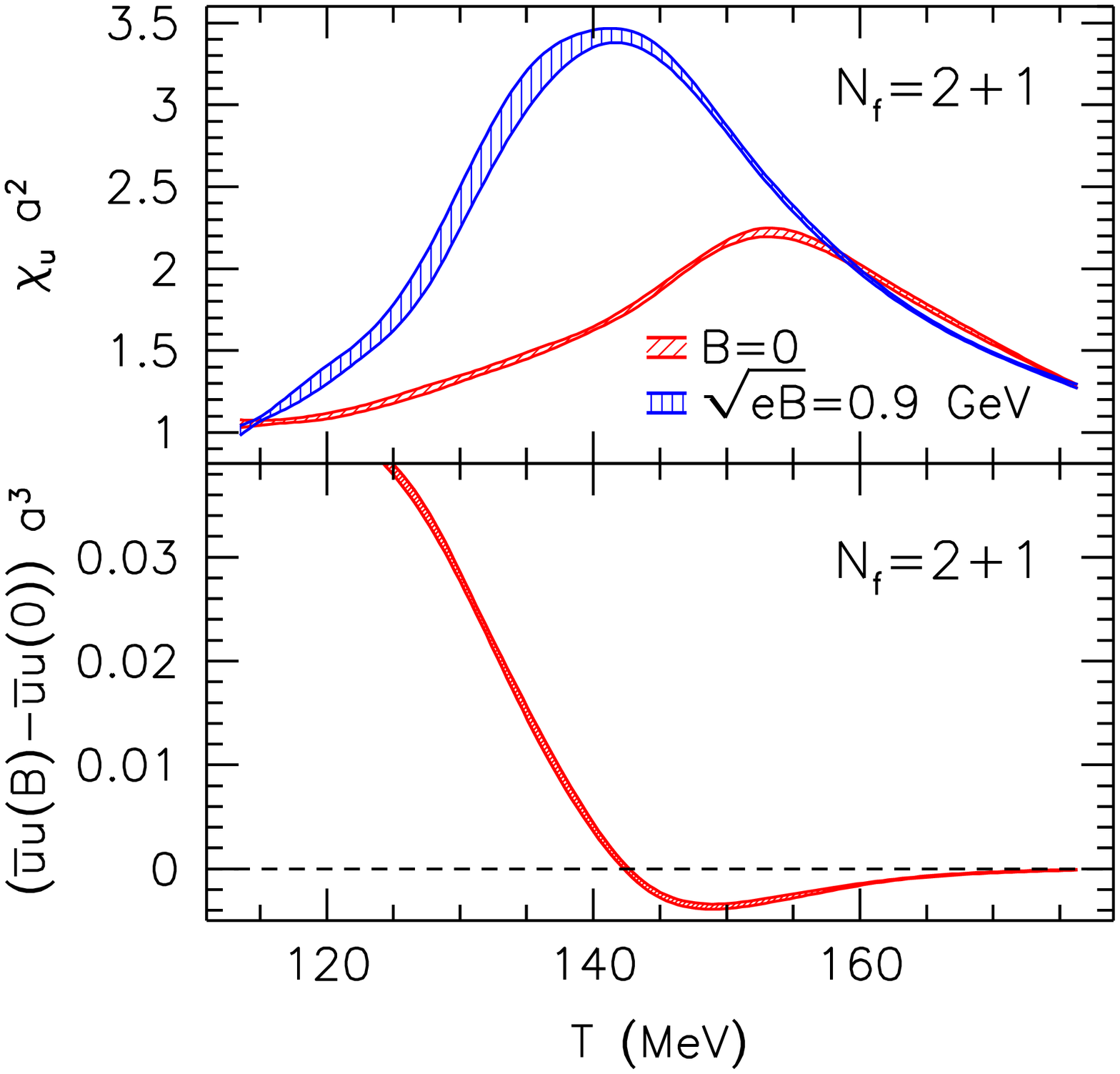}\quad
\includegraphics*[width=7.0cm]{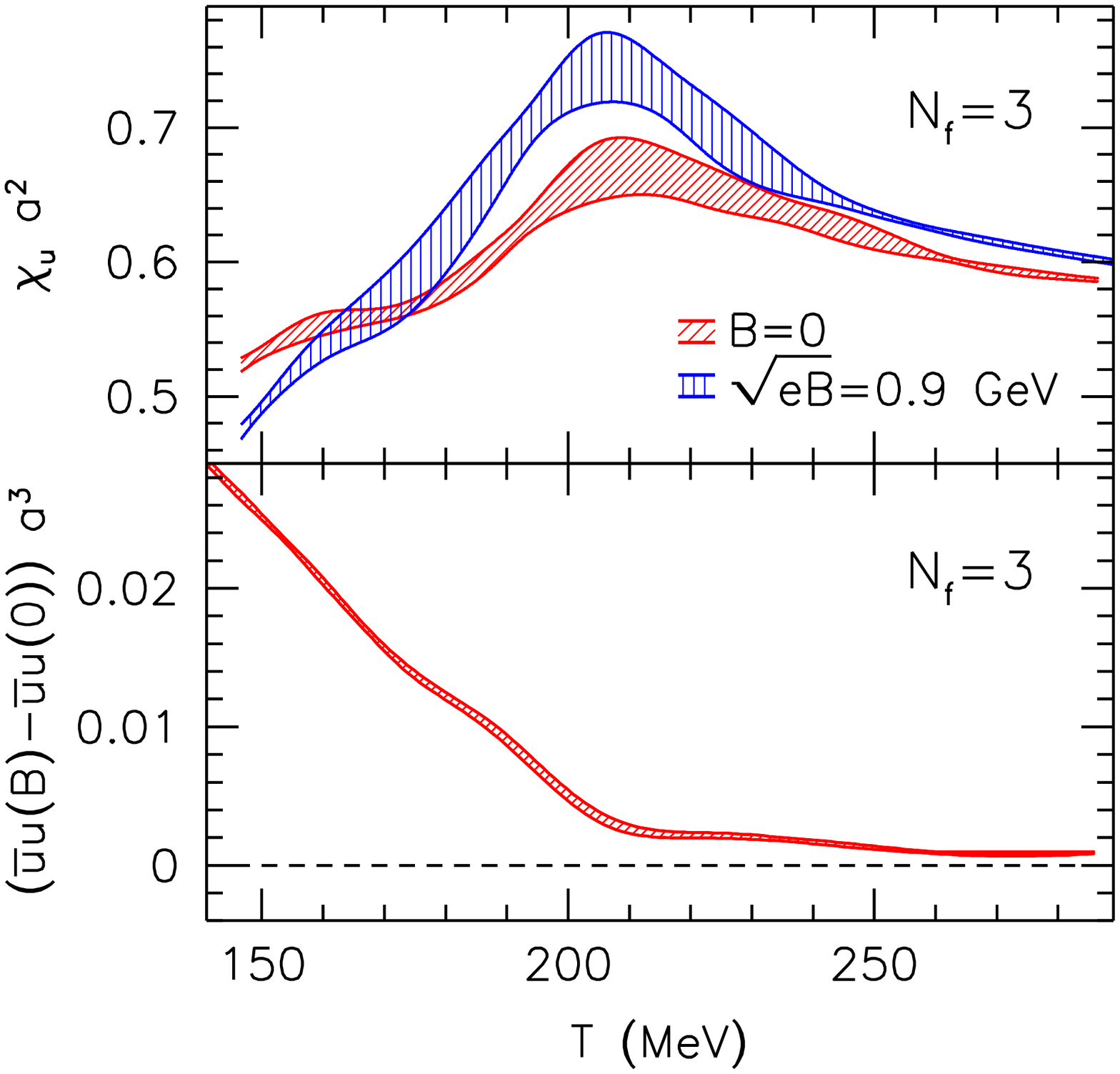}
}
\vspace*{-0.4cm}
\caption{The bare chiral susceptibility for large and vanishing magnetic fields (upper panels), and the difference between the up condensates at $\sqrt{eB}=0.9$ GeV and at $B=0$ (lower panels) for the $N_t=6$ lattices. Results are shown for the $N_f=2+1$ theory (left panels) and for the $N_f=3$ theory, where each quark has the physical strange quark mass (right panels).}
\label{fig:threeflavor}
\end{figure}

We summarize our findings as a) the dependence of the condensate on the external field is non-monotonic and varies strongly with temperature, and b) as a result the pseudocritical temperature shifts to lower values at large $B$ as compared to the $B=0$ case. The latter observation is supported by a similar $T_c(B)$ dependence deduced from the chiral susceptibility or the strange quark number susceptibility, see section~\ref{sec:phasediagram}. We investigate the reason for this behavior further by increasing our light quark masses up to the physical strange quark mass, studying the $N_f=3$ theory. As a first approximation we apply the same lattice scale and line of constant physics as was used for the $N_f=2+1$ flavor analysis. This clearly introduces a systematic error, but most probably does not affect the qualitative behavior. In figure~\ref{fig:threeflavor} we show the $N_f=2+1$ results for the chiral condensate and susceptibility (left panels), compared to the $N_f=3$ data (right panels). In the upper panels we plot the unrenormalized chiral susceptibility for the case of a vanishing external field (red bands) and a large field of $\sqrt{eB}=0.9$ GeV (blue bands). Furthermore, in order to see the change in the condensate due to the presence of the external field, we plot the difference between the condensate at $\sqrt{eB}=0.9$ GeV and at $B=0$ in the lower panels. For the case of $N_f=2+1$ we plot the $B=\textmd{const.}$ slice of the 2-dimensional surfaces we obtained as described in section~\ref{sec:analysis}, while for $N_f=3$ we fit the data to a simple spline function (in the latter analysis we keep the physical value of the magnetic field constant by tuning $N_b\sim B/T^2$ as a function of $T$ to keep $B$ fixed).

As is clearly visible in the lower left panel of figure~\ref{fig:threeflavor}, the magnetic field reduces the chiral condensate for temperatures $T\gtrsim 140$ MeV, thus pushing the inflection point of the condensate towards the left and causing a decrease in $T_c(B)$. This decrease is also visible from the behavior of the corresponding susceptibility, shown in the upper left panel of the figure. On the other hand, for larger quark masses the situation drastically changes: the condensate increases with the magnetic field for all temperatures (see the lower right panel of figure~\ref{fig:threeflavor}), similarly as was observed in~\cite{D'Elia:2010nq}. Moreover, there is no clear change in $T_c$:
the chiral susceptibility (upper right panel) gives consistent pseudocritical temperatures for both $B=0$ and $\sqrt{eB}=0.9$ GeV. This observation
supports our explanation number three, namely that the difference regarding the change in $T_c(B)$ between the present work and the study of~\cite{D'Elia:2010nq} stems at least partially from the larger-than-physical quark masses of the latter.\footnote{Note that while the mass of the Goldstone pion of~\cite{D'Elia:2010nq} is below 200 MeV, due to the larger taste splitting the higher lying pion tastes may have a larger impact on the response to the magnetic field.}

\begin{wrapfigure}{r}{7.0cm}
\centering
\vspace*{-0.2cm}
\includegraphics*[width=6.8cm]{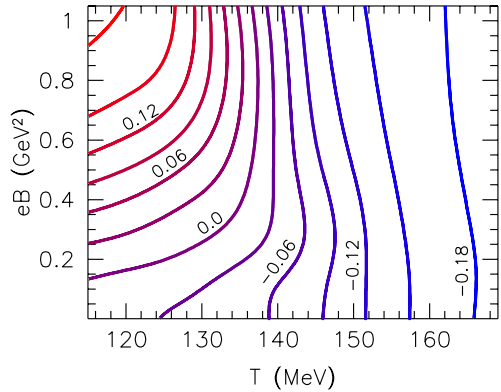}
\vspace*{-0.2cm}
\caption{Contour plot of $\bar{u}u^r$.}
\label{fig:pbpcontours}
\vspace*{-0.5cm}
\end{wrapfigure}
To illustrate the behavior of the condensate from yet another aspect, we show in figure~\ref{fig:pbpcontours} the contour plot of the renormalized chiral condensate as a function of $T$ and $B$. The color of the curves encodes the value of the condensate, ranging from $-0.18$ to $0.18$ (blue towards red) in steps of $0.03$. A similar plot of the results of~\cite{D'Elia:2010nq} would consist of curves having positive slopes, indicating that each point of $\bar u u$ moves towards the right as a result of a finite $B$. Here we find that for example the $\bar{u}u^r=-0.12$ curve clearly has segments with negative slope, which once again reflects the complex behavior of the condensate as a function of $B$ and $T$.

\section{Nature of the transition - finite size effects}
\label{sec:finitesize}

Here we address the question of how the strength of the transition changes as the external field is switched on.
At $B=0$ the transition is known to be a broad crossover~\cite{Aoki:2006we}, where the approximate order parameters like the chiral condensate change smoothly with the temperature, and no finite volume scaling is visible in the observables. 
Furthermore, the crossover nature of the transition implies that -- as we will also observe, see figure~\ref{fig:phasediag_cont} -- different observables give different pseudocritical temperatures~\cite{Aoki:2006br}.

\begin{figure}[h!]
\centering
\vspace*{-0.3cm}
\mbox{
\includegraphics*[width=5.7cm]{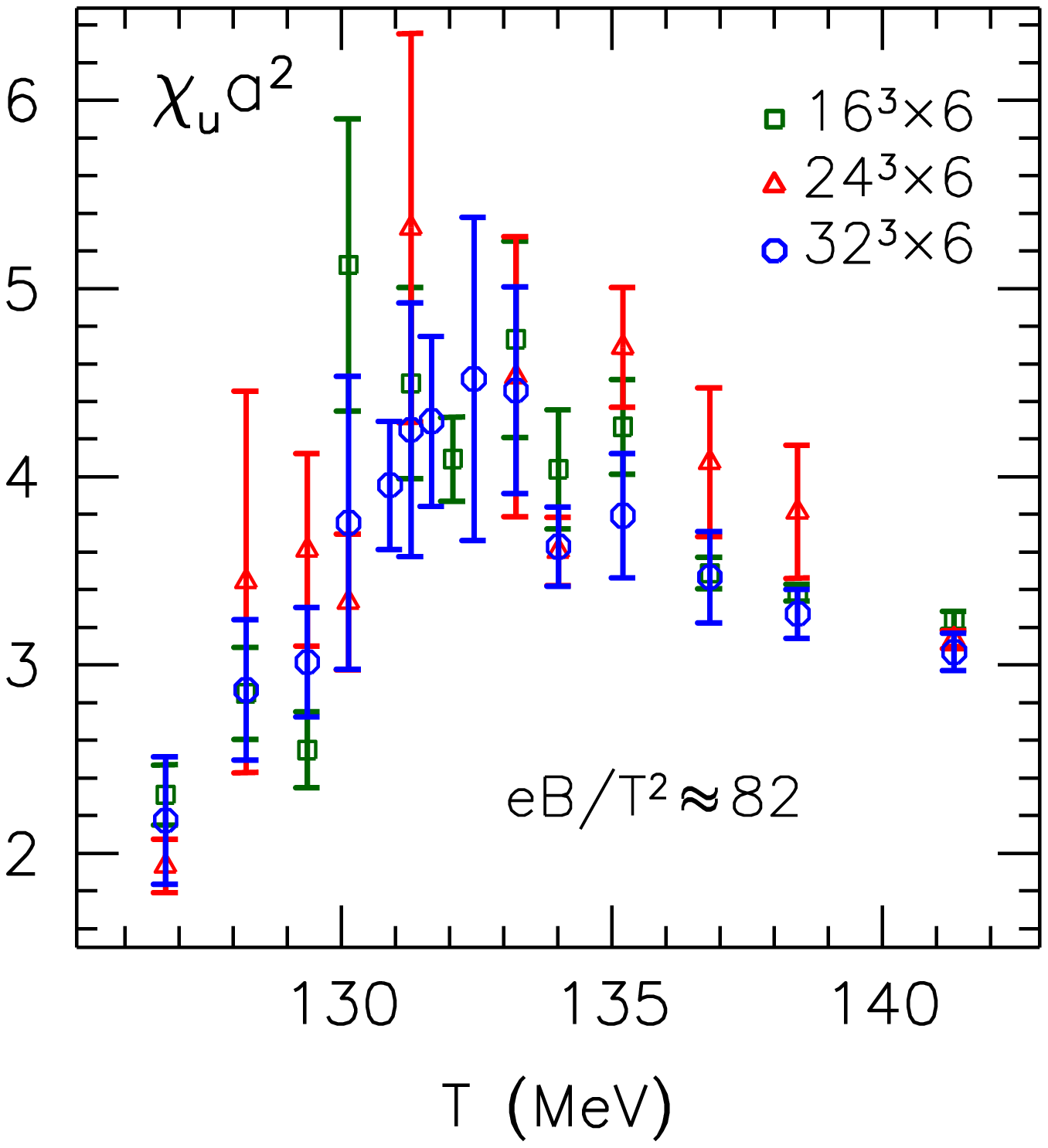} \hspace*{0.7cm}
\includegraphics*[width=5.7cm]{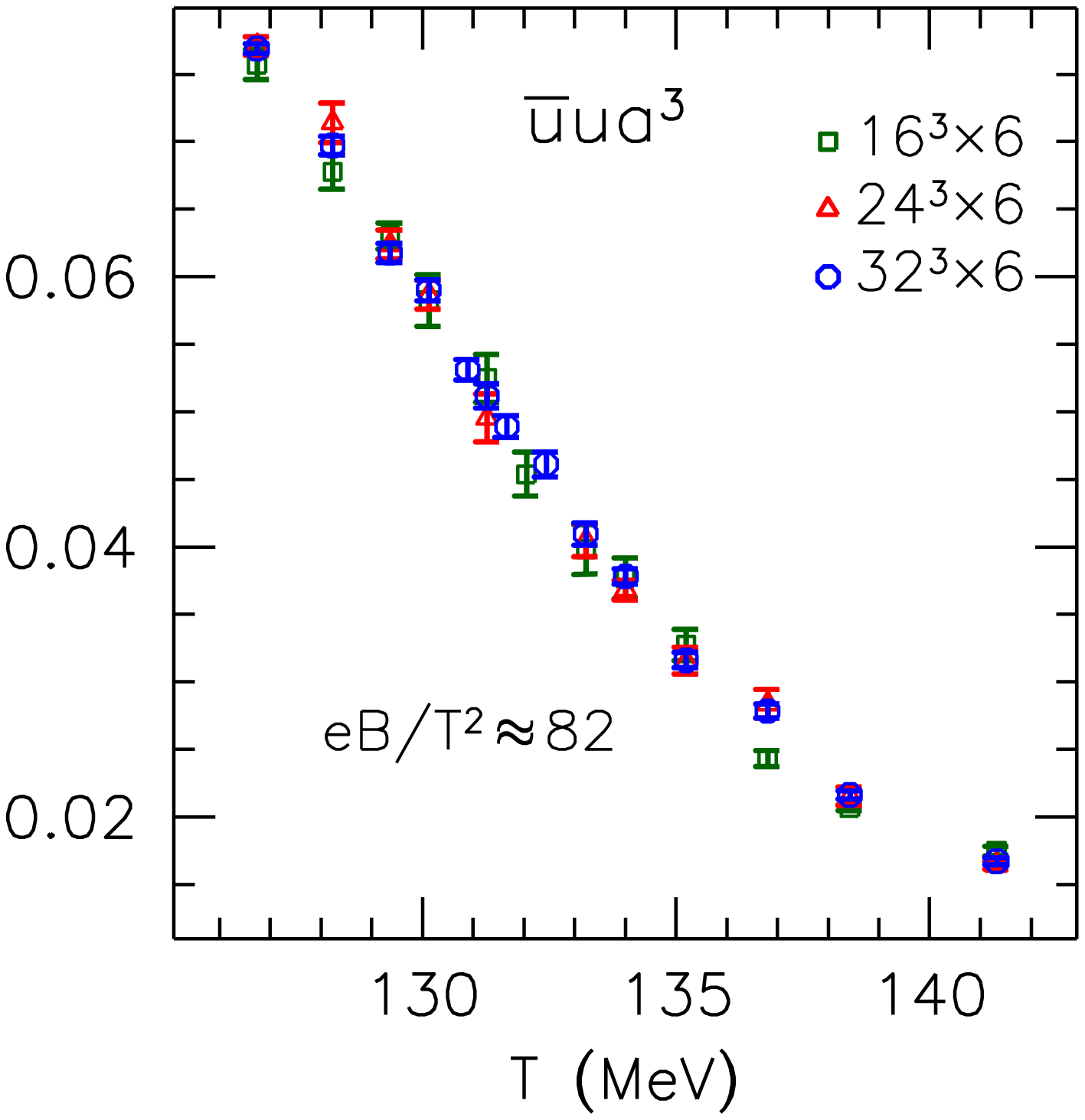}
}
\caption{The unrenormalized chiral susceptibility (left panel) and chiral condensate (right panel) as functions of $T$ measured on our $N_t=6$ lattices for different spatial volumes. No finite size effects are visible within statistical errors. The crossover nature of the transition persists up to this large external field.}
\label{fig:voldep}
\end{figure}

As can be seen from figure~\ref{fig:surface}, not just the transition temperature changes with $B$, but also the shapes of our observables as functions of $T$ are altered by a finite magnetic field. 
Specifically, we find that the maximum value of the chiral susceptibility $\chi_u^r$ increases with $B$, which may suggest the transition to become stronger for large magnetic fields, as was also reported in~\cite{D'Elia:2010nq}. 
To properly determine the nature of the transition we search for finite volume scaling in our observables.

To this end we perform simulations at our largest magnetic field on the $N_t=6$ lattices with $N_s=16,24$ and $32$. The largest lattice in the transition region corresponds to a box of linear size $\sim 7$ fm. 
Here we keep $eB/T^2$ fixed (and not $B$ itself) as we are only interested in differences between the various volumes. In figure~\ref{fig:voldep} the results for the chiral susceptibility (left panel) and for the chiral condensate (right panel) are shown as functions of the temperature for $eB/T^2\approx82$. 
The figure shows that our $N_s=16$ results agree within statistical errors with the $N_s=24$ and $N_s=32$ data, indicating that finite size errors are small, compared to statistical errors. This observation also implies that the transition at this high magnetic field is still an analytic crossover. 

\begin{wrapfigure}{r}{8.0cm}
\centering
\vspace*{-0.6cm}
\includegraphics*[width=7.5cm]{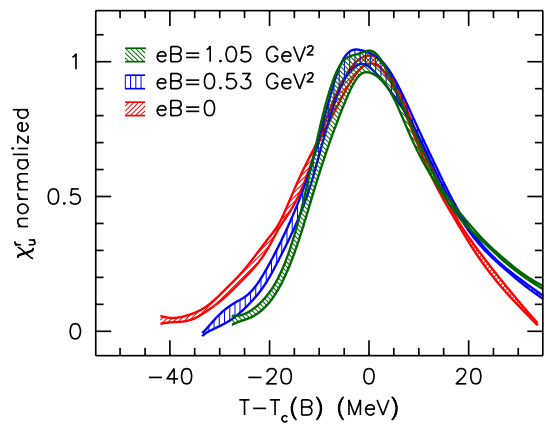}	
\vspace*{-0.1cm}
\caption{Relative changes in the $T$-dependence of $\chi_u^r$ as measured on $N_t=6$ lattices. The width of the peak decreases only mildly.}
\label{fig:chi_width}
\vspace*{-0.4cm}
\end{wrapfigure} 

To further study how the strength of the transition changes we investigate the width of the renormalized chiral susceptibility. In figure~\ref{fig:chi_width} we plot the susceptibility divided by its maximum value 
as a function of $T-T_c(B)$ for three different values of the magnetic field for the $N_t=6$ lattices. 
We find that, although the height of the peak in $\chi_u^r$ grows significantly (by almost a factor of 2 between $B=0$ and the largest $B$, see also upper left panel of figure~\ref{fig:threeflavor}), the width of the peak is only mildly affected by the magnetic field. In particular, the width of the peak at half maximum decreases from $\sim 30(3)$ MeV to $\sim 25(3)$ MeV as the external field is increased from zero to $eB=1.05 \textmd{ GeV}^2$. 
We find a very similar behavior on the $N_t=8$ and $10$ lattices.
From this analysis our final conclusions are that the width of the transition decreases only mildly with increasing magnetic field, and as the finite size scaling analysis has shown, the transition remains an analytic crossover at least up to $\sqrt{eB}\sim 1$ GeV.

\section{The phase diagram}
\label{sec:phasediagram}

\begin{figure}[h!]
\centering
\vspace*{-0.4cm}
\mbox{
\includegraphics*[width=7.0cm]{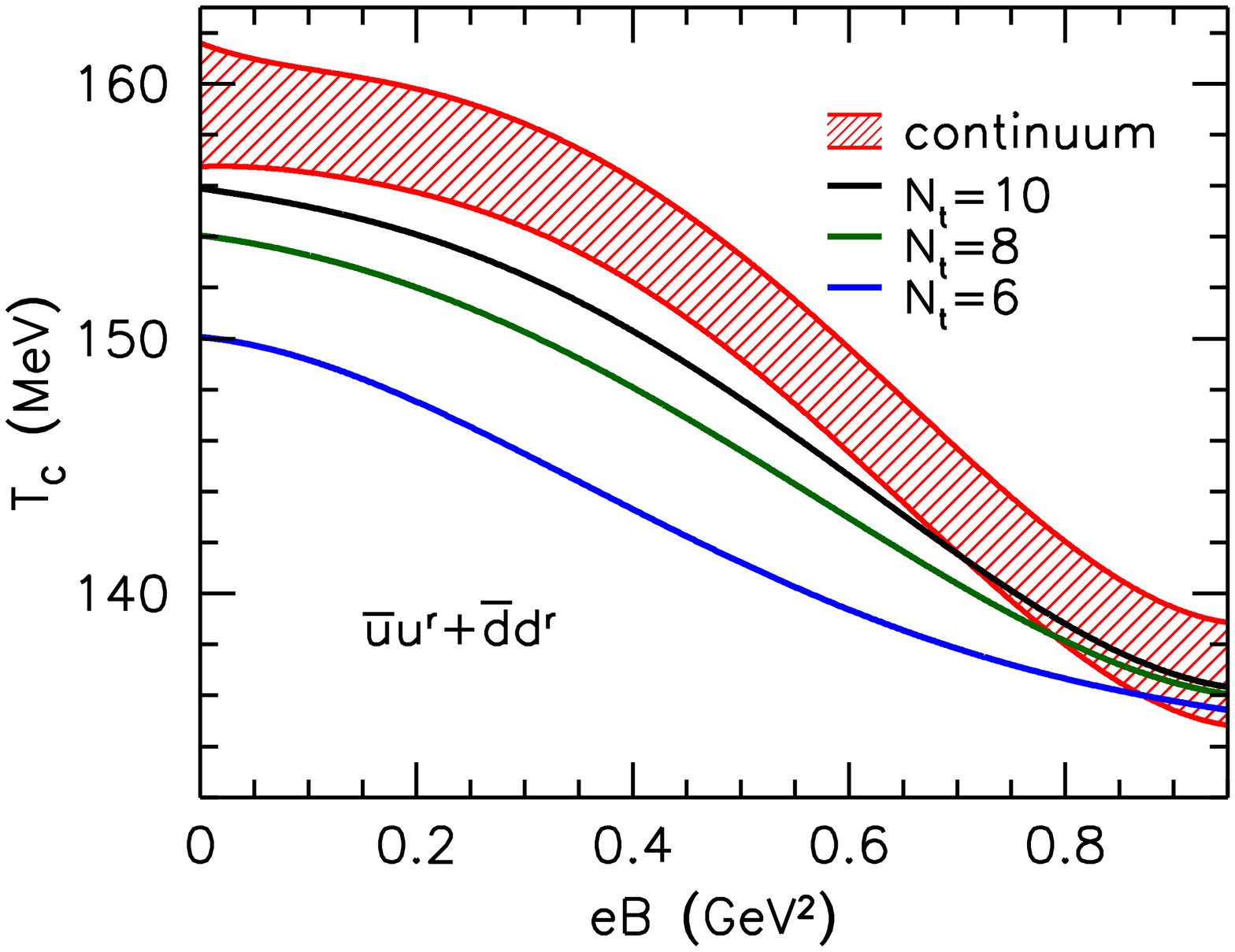}\quad
\includegraphics*[width=7.0cm]{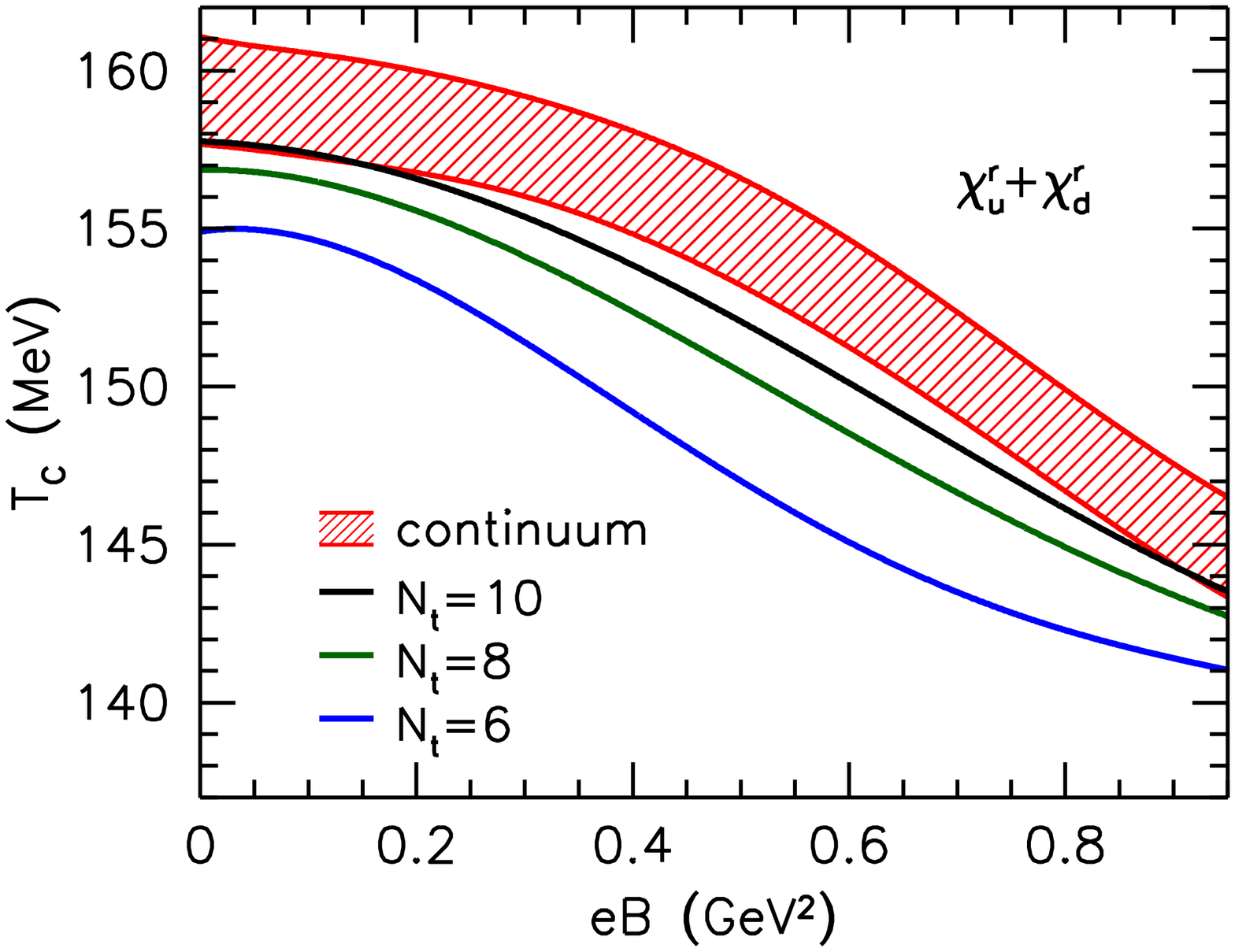} }\vspace*{-0.3cm}
\includegraphics*[width=7.0cm]{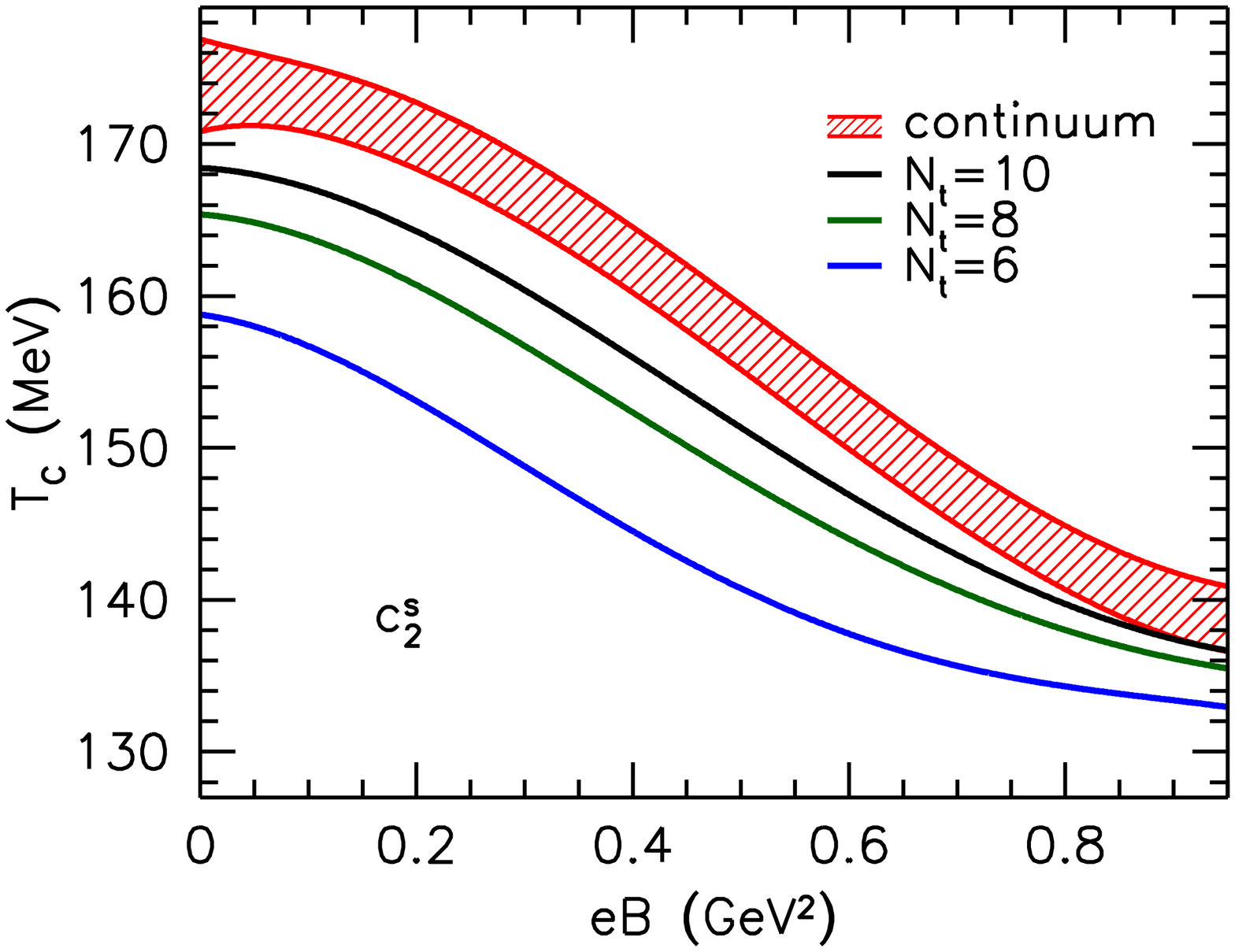}
\vspace*{-0.1cm}
\caption{The phase diagram of QCD in the $B-T$ plane, determined from the renormalized chiral condensate $\bar u u^r + \bar d d^r$ (upper left panel), the renormalized chiral susceptibility $\chi_u^r+\chi_d^r$ (upper right) and the strange quark number susceptibility $c_2^s$ (lower panel).}
\label{fig:phasediag_cont}
\end{figure}

Finally, using the fitted two-dimensional surfaces of section~\ref{sec:analysis}, we study the observables as functions of the temperature, along the lines of constant magnetic field. 
In particular we analyze the renormalized chiral susceptibility $\chi_u^r + \chi_d^r$, the renormalized chiral condensate $\bar u u^r + \bar d d^r$ and the strange quark number susceptibility $c_2^s$. For the latter two observables we determine the pseudocritical temperature $T_c(B)$ as the inflection points of the curves, while for the former we calculate the position of the maximum value of the observable. The results are shown in figure~\ref{fig:phasediag_cont}. 

To carry out the continuum extrapolation, we fit the results for $T_c(B)$ for all three lattice spacings ($N_t=6,8$ and $10$) together with an $N_t$-dependent polynomial function of order four of the form $T_c(B,N_t) = \sum_{i=0}^4(a_i + b_i N_t^{-2}) B^i$.
This ensures the scaling of the final results with $N_t^{-2}\sim a^2$. We obtain $\chi^2/{\rm dof.}\approx 0.5 \ldots 1.2$ indicating good fit qualities. 
In order not to make the plots overcrowded, we only show error bars for the continuum curves. The error coming from the continuum extrapolation is estimated to be 2 MeV and is added to the statistical error in quadrature. The error in the lattice scale determination~\cite{Borsanyi:2010cj} propagates in the $T_c(B)$ function and amounts to an additional $2-3$ MeV systematic error, which is not added to the errors for figure~\ref{fig:phasediag_cont} since we find that it does not influence the shape of the curves.

As is clearly visible in figure~\ref{fig:phasediag_cont}, all three observables show that the pseudocritical temperature {\it decreases} with growing external field $B$. Preliminary results for the Polyakov loop at one lattice spacing show a very similar decrease in $T_c(B)$, see appendix~\ref{app:Poly}.
We observe that the strange susceptibility (which can be viewed as a quantity signaling the deconfinement transition) is the observable most sensitive to the external field. $T_c(B)$ changes most drastically in this case, by almost $35$ MeV between $B=0$ and $eB\approx 1 \textmd{ GeV}^2$. We note that our results at $B=0$ are all consistent with earlier determinations of the pseudocritical temperature where the stout smeared staggered lattice action was used~\cite{Aoki:2006br,Aoki:2009sc,Borsanyi:2010bp}. 
We mention that one would expect $\mathcal{O}(a^2)$ effects to become more pronounced as the magnetic field grows. However, the numerical data of the pseudocritical temperatures seem to scale well, even up to our maximum value of $eB\approx 1 \textmd{ GeV}^2$.

\section{Summary}

In this paper we studied the finite temperature transition of QCD in the presence of external (electro)magnetic fields via lattice simulations at physical quark masses. The extrapolation to the continuum limit is carried out, and finite size effects are under control. The results are relevant for the description of both the evolution of the early universe and of noncentral heavy ion collisions. 

We obtained the phase diagram of QCD in the $B-T$ plane using three observables in the phenomenologically interesting region of $0\le eB \lesssim 1 \textmd{ GeV}^2$. 
Performing a finite volume scaling study we found that the transition remains an analytic crossover up to our largest magnetic fields, with the transition width decreasing only mildly.
This rules out the existence of a critical endpoint in the $B-T$ phase diagram below $eB=1 \textmd{ GeV}^2$.
Moreover, our results indicate that the transition temperature significantly {\it decreases} with increasing $B$. This result contradicts several model calculations present in the literature which predict an increase in $T_c$ as $B$ grows (see the summary in section~\ref{sec:intro}). We presented indications that the response of $T_c$ to the external field can be traced back to the behavior of the chiral condensate as a function of $T$ and $B$. We showed that this behavior is more complex than is predicted by most model calculations (where the condensate increases with $B$ for any temperature), and that it depends very strongly on the quark masses.

\begin{figure}[ht!]
\centering
\vspace*{-0.4cm}
\includegraphics*[width=9.5cm]{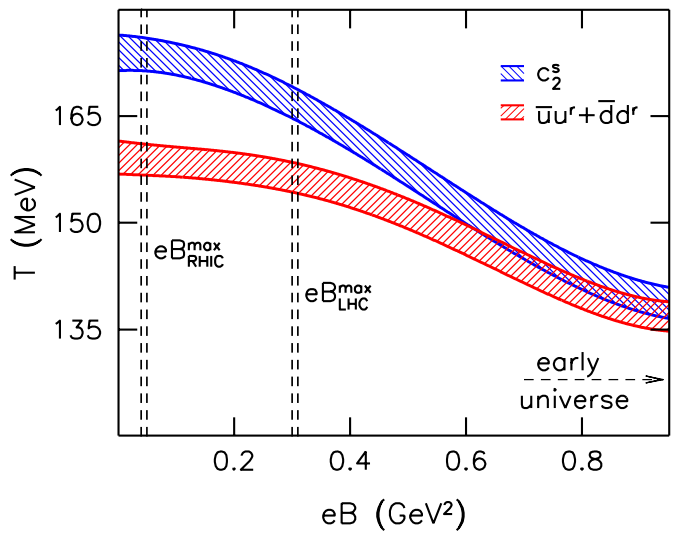}
\caption{Our final result: the QCD phase diagram in the magnetic field - temperature plane. The colored bands represent the pseudocritical temperature as defined from inflection points of the renormalized chiral condensate $\bar u u^r + \bar d d^r$ (red) and the strange quark number susceptibility $c_2^s$ (blue) in the continuum limit. Also indicated by the dashed vertical lines are the maximal magnetic fields produced at RHIC and at the LHC. The large $B$ region of the phase diagram is relevant for the evolution of the early universe.
 }
\label{fig:finalres}
\end{figure}

We summarize our results in figure~\ref{fig:finalres}, which shows the QCD phase diagram in the $B-T$ plane as defined using the renormalized chiral condensate $\bar u u^r +\bar d d^r$ and the strange quark number susceptibility $c_2^s$ in the continuum limit. By comparing our magnetic fields to the maximal fields that may be produced in noncentral heavy ion collisions we conclude that the decrease in $T_c$ is negligible for RHIC and may be up to $5-10$ MeV for the LHC. Moreover, the effect grows with the magnetic field, exceeding 20$\%$ for $c_2^s$ at $eB=1 \textmd{ GeV}^2$. This may have a significant impact on the description of the QCD transition during the evolution of the early universe.

\acknowledgments{

This work has been supported by DFG grants SFB-TR 55, FO 502/1-2 and BR 2872/4-2, the EU grants (FP7/2007-2013)/ERC no 208740 and PITN-GA-2009-238353 (ITN STRONGnet). Computations were carried out on the GPU cluster at the E\"otv\"os University in Budapest and on the Bluegene/P at FZ J\"ulich. We thank Ferenc Niedermayer for useful discussions, interesting ideas and for careful reading of the manuscript.
G.~E. would like to thank Massimo D'Elia, Swagato Mukherjee, D\'aniel N\'ogr\'adi, Tam\'as Kov\'acs and Igor Shovkovy for useful discussions. 
}

\appendix
\section{Lattice vector potential and periodic boundary conditions}
\label{sec:app1}

In this appendix we show that the lattice prescription for the $\mathrm{U}(1)$ links, as in equation~(\ref{eq:links2}), is indeed equivalent to the continuum vector potential up to a local $\mathrm{U}(1)$ gauge transformation.
The direct lattice discretized version of the continuum vector potential~(\ref{eq:contvecpot}) can be written as
\be
\begin{split}
u_\nu(n) &=1, \quad \quad\quad\quad (\nu \ne y), \\
u_y(n) &= e^{i a^2 q B n_x}.
\end{split}
\label{eq:links1}
\ee
The periodic boundary conditions are now only satisfied up to a local $\mathrm{U}(1)$ gauge transformation (transition function),
\be
\begin{split}
u_y(N_x,n_y,n_z,n_t) &= u_y(0,n_y,n_z,n_t) \cdot V, \\
V &= e^{i a^2 qB N_x},
\end{split}
\ee
where $u$ is the abelian gauge field and $V$ the gauge transformation that acts in $\mathrm{U}(1)$ space. However, on the lattice it is more convenient to have exactly periodic boundary conditions. Hence we perform the inverse $\mathrm{U}(1)$ gauge transformation on the last $x$-slice of the lattice, which changes fermions $\psi$ as
\be
\psi(N_x,n_y,n_z,n_t) \to \psi(N_x,n_y,n_z,n_t) \cdot V^{n_y},
\ee
and the links in both the $x$ and $y$ directions as
\be
\begin{split}
u_y(N_x,n_y,n_z,n_t) &\to u_y(N_x,n_y,n_z,n_t) \cdot V^{-1}, \\
u_x(N_x-1,n_y,n_z,n_t) &\to  u_x(N_x-1,n_y,n_z,n_t) \cdot V^{-n_y},
\end{split}
\ee
resulting in periodic boundary conditions and the ``twisted'' links that we presented in equation~(\ref{eq:links2}).

\section{Renormalization properties of \boldmath \texorpdfstring{$eB$}{eB} from \texorpdfstring{$\mathrm{U}(1)$}{u} gauge invariance}
\label{app:WT}

QCD with an external magnetic field has a local $\mathrm{U}(1)$ gauge invariance. Let $\psi_0$ be the bare quark field and $A_0$ and $e_0$ the bare external electromagnetic field and electromagnetic coupling. The renormalization of these are given as,
\be
\psi^R=\sqrt{Z_2}\cdot \psi^0, \quad\quad A_\mu^R=\sqrt{Z_3} \cdot A_\mu^0, \quad\quad e^R=Z_e \cdot e^0.
\ee
If both the regularization and the renormalization prescriptions are gauge invariant, then so are the renormalization constants $Z_2, Z_3$ and $Z_e$. The gauge transformation for the bare and renormalized quark fields is of the form
\be
{\psi^0}' = \psi^0 \exp\left(i\alpha\right), \quad\quad {\psi^R}' = \psi^R \exp\left(i\alpha^R\right),
\ee
which shows that $\alpha=\alpha^R$, due to the gauge invariance of $Z_2$. The same transformations for the external electromagnetic field are then
\be
{A_\mu^0}' = A_\mu^0 + \frac{1}{e^0} \partial_\mu \alpha, \quad\quad
{A_\mu^R}' = A_\mu^R + \frac{1}{e^R} \partial_\mu \alpha.
\label{eq:bareA}
\ee
Dividing the second equation by $\sqrt{Z_3}$ and equating it to the first we obtain (inserting $e^R=Z_e\cdot e^0$),
\be
A_\mu^0 + \frac{1}{Z_e\sqrt{Z_3}\cdot e^0} \partial_\mu \alpha = A_\mu^0 + \frac{1}{e^0} \partial_\mu \alpha,
\ee
which implies
\be
Z_e \sqrt{Z_3} = 1.
\ee
This is a well known result in QED which therefore also applies for the case of QCD with an external magnetic field. Since $Z_e\sqrt{Z_3}$ is the particular combination which renormalizes the product $eB$, our conclusion is that $eB$ does not need renormalization.

\section{Simulation parameters}
\label{sec:simpar}

In this appendix we tabulate the simulation parameters for the $T=0$ and the $T>0$ runs. The number of thermalized trajectories generated for each set of parameters $(\beta, N_b)$ ranges from several hundred to a few thousand.

\begin{table}[ht!]
\begin{center}
\begin{tabular}{| c | c | c |}
\hline
\multirow{3}{*}{$16^3\times 6$} & \multirow{2}{*}{$\beta$} & 3.45, 3.465, 3.48, 3.488, 3.492, 3.495, 3.497, 3.5, \\ 
& & 3.505, 3.507, 3.51, 3.514, 3.518, 3.525, 3.54 \\ \cline{2-3}
& $N_b$ & 31 \\
\hline \hline
\multirow{3}{*}{$24^3\times 6$} & \multirow{2}{*}{$\beta$} & 3.45, 3.465, 3.48, 3.495, 3.51, 3.525, \\
& & 3.54, 3.555, 3.57, 3.585, 3.6, 3.625 \\ \cline{2-3}
& $N_b$ & 0, 1, 2, 3, 4, 8, 16, 24, 32, 50, 70 \\
\hline \hline
\multirow{3}{*}{$32^3\times 6$} & \multirow{2}{*}{$\beta$} & 3.465, 3.48, 3.488, 3.492,  3.495, 3.497, 3.5, 3.503, \\
& & 3.505, 3.507, 3.51, 3.514, 3.518, 3.525, 3.54 \\ \cline{2-3}
& $N_b$ & 124 \\
\hline \hline
\multirow{3}{*}{$24^3\times 8$} & \multirow{2}{*}{$\beta$} & 3.525, 3.55, 3.575, 3.6, 3.625, 3.64,  \\
& & 3.65, 3.675, 3.7, 3.725, 3.75, 3.775 \\ \cline{2-3}
& $N_b$ & 0, 1, 2, 3, 4, 5, 7, 9, 13, 18, 29, 40 \\
\hline \hline
\multirow{3}{*}{$28^3\times 10$} & \multirow{2}{*}{$\beta$} & 3.6, 3.625, 3.65, 3.675, 3.687, 3.7, 3.712, 3.725,\\
& & 3.738, 3.75, 3.762, 3.775, 3.8, 3.825, 3.85, 3.875 \\ \cline{2-3}
& $N_b$ & 0, 1, 2, 4, 8, 12, 16, 25, 34, 40 \\
\hline 
\end{tabular}

\caption{Simulation points for the $T>0$ runs.}
\end{center}
\end{table}

\begin{table}[ht!]
\begin{center}
\begin{tabular}{| c | c | c |}
\hline
\multirow{2}{*}{$24^3\times 32$} & $\beta$ & 3.45, 3.55 \\ \cline{2-3}
& $N_b$ & 1, 2, 3, 4, 8, 12, 16, 20, 24, 32, 70 \\
\hline \hline 
\multirow{2}{*}{$32^3\times 48$} & $\beta$ & 3.67 \\ \cline{2-3}
& $N_b$ & 1, 2, 3, 4, 5, 6, 12, 24 \\
\hline \hline
\multirow{2}{*}{$40^3\times 48$} & $\beta$ & 3.75 \\ \cline{2-3}
& $N_b$ & 1, 2, 3, 4, 6, 12 \\
\hline
\end{tabular}

\caption{Simulation points for the $T=0$ runs.}
\end{center}
\end{table}

\section{Code check}
\label{sec:codecheck}

\begin{wrapfigure}{r}{8.8cm}
\centering
\vspace*{-0.3cm}
\includegraphics*[width=8cm]{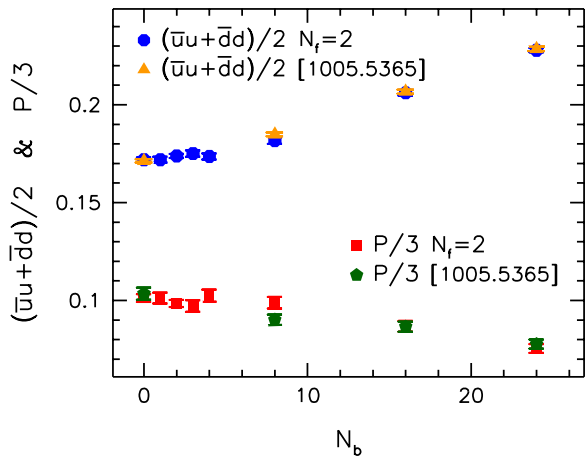}
\vspace*{-0.2cm}
\caption{The average condensate and the Polyakov loop as functions of the magnetic flux in the $N_f=2$ theory, compared to results of~\cite{D'Elia:2010nq}.}
\vspace*{-0.2cm}
\label{fig:check}
\end{wrapfigure}

To check our code and simulation algorithm we reproduced the results of~\cite{D'Elia:2010nq} at one temperature. We employ exactly the same simulation setup, i.e. we use the Wilson gauge action and $N_f=2$ flavors of unsmeared naive staggered quarks on a $16^3\times 4$ lattice. We measure the up and down quark condensates and the Polyakov loop (see definition in appendix~\ref{app:Poly}) at gauge coupling $\beta=5.35$ and mass $am=0.075$. We plot the average condensate and the Polyakov loop in figure~\ref{fig:check} (to conform to the notation of~\cite{D'Elia:2010nq} we divide our Polyakov loops by 3).
We observe that results for both $\bar{u}u+\bar{d}d$ and $P$ agree within statistical errors, with the exception of one point for $P$ where the values differ by $2\sigma$, as expected for 8 points on statistical grounds. Therefore we confirm that there is no discrepancy between results from the two algorithms/implementations.

\section{Polyakov loop}
\label{app:Poly}

We carry out the same analysis as presented in section~\ref{sec:analysis} for the Polyakov loop,
\be
P \equiv \frac{1}{V}\! \sum\limits_{n_x,n_y,n_z}\! \Tr \prod\limits_{n_t=0}^{N_t-1} U_4(n).
\ee
We note that while the quark condensates and susceptibilities and the quark number susceptibility depend explicitly on the magnetic field, the Polyakov loop, as a purely gluonic operator, is only affected by the modified
spatial links indirectly; its expectation value at $B>0$ is influenced by the magnetic factors of equation~(\ref{eq:links2}) appearing in the fermion determinant.
To cancel the multiplicative divergences of $P$, we define the renormalized
Polyakov loop~\cite{Aoki:2006br} using the static quark-antiquark potential $V(r)$
\begin{wrapfigure}{r}{8.8cm}
\centering
\vspace*{-0.1cm}
\includegraphics*[height=6.2cm]{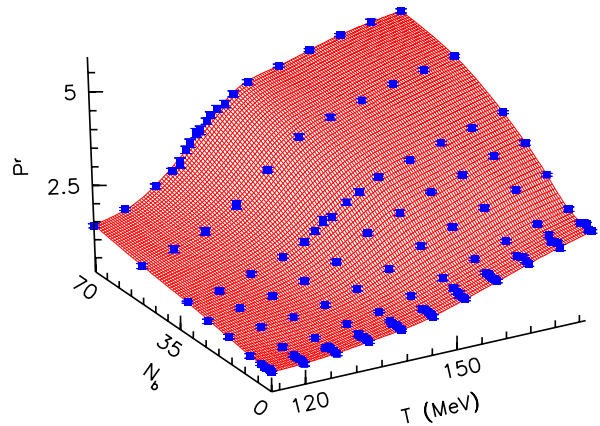}
\vspace*{-0.5cm}
\caption{The renormalized Polyakov loop as a function of $T$ and $N_b$ on the $24^3\times 6$ lattices. Measurements are denoted by the blue points, while the red surface is the spline fit to the data.}
\label{fig:Polyakov}
\vspace*{-3cm}
\end{wrapfigure}
as
\be
P^r(B,T) = P(B,T) e^{V(r_0)/2T}.
\ee
We measure this observable and perform the spline fitting, see figure~\ref{fig:Polyakov} for the $N_t=6$ results. 
The inflection point moves to smaller temperatures as we increase the
magnetic field. This behaviour is similar to the decreasing transition
temperature observed for the condensate or susceptibility (c.f. figure~\ref{fig:surface}).

\bigskip\bigskip\bigskip\bigskip\bigskip

\bibliographystyle{jhep}
\bibliography{magnetic}

\end{document}